\PassOptionsToPackage{table}{xcolor}
\documentclass[a4paper,fleqn]{cas-dc}  

\usepackage{amsmath,amssymb,amsfonts}

\usepackage{algorithmic}

\usepackage{graphicx}
\usepackage{placeins}          
\usepackage{subcaption}        
\usepackage{wrapfig}
\usepackage[most]{tcolorbox}

\usepackage{tabularx}
\usepackage{makecell}
\usepackage{multirow}
\usepackage{float}
\usepackage{longtable}
\usepackage{placeins}
\usepackage{array}

\newcolumntype{Y}{>{\raggedright\arraybackslash}X} 

\usepackage{colortbl}
\definecolor{sectiongray}{RGB}{191,191,191}

\usepackage{textcomp}
\usepackage{balance}
\usepackage{enumitem}
\usepackage{soul}
\usepackage{fontawesome}

\usepackage{listings}

\usepackage{tikz}
\usepackage{pgfplots}
\pgfplotsset{compat=1.7}

\usepackage{rotating}
\usepackage{pdflscape}

\usepackage[numbers]{natbib}

\usepackage{hyperref}
\hypersetup{
  colorlinks = true,
  urlcolor   = blue,
  linkcolor  = blue,
  citecolor  = blue
}

\shorttitle{Quantum Computing as a Service (QCaaS)}
\shortauthors{Ahmad et al.}

\title[mode=title]{Quantum Computing as a Service -- a Software Engineering Perspective\tnotemark[1]}

\author[LUL]{Aakash Ahmad}
\ead{a.ahmad13@lancaster.ac.uk}
\credit{Conceptualisation, Methodology, Investigation, Data curation, Software, Writing - Original draft preparation}

\author[JYU]{Muhammad Waseem}
\ead{mwaseem@jyu.fi}

\author[UQU]{Bakheet Aljedaani}
\ead{bhjedaani@uqu.edu.sa}

\author[USQ]{Mahdi Fahmideh}
\ead{mahdi.fahmideh@usq.edu.au}

\author[WHU]{Peng Liang}
\ead{liangp@whu.edu.cn}

 \author[UMU]{Feras Awaysheh}
 \ead{feras.awaysheh@umu.se}

\affiliation[LUL]{organization={University of Derby, School of Computing, College of Science and Engineering},
  city={Derby}, country={United Kingdom}}

\affiliation[JYU]{organization={Faculty of Information Technology and Communication Sciences, Tampere University},
  city={Tampere}, country={Finland}}

  \affiliation[UQU]{organization={Computer Science Department, Aljumum University College, Umm Alqura University},
  city={Makkah}, country={Saudi Arabia}}

  \affiliation[USQ]{organization={School of Business, University of Southern Queensland},
  city={Queensland}, country={Australia}}
  
\affiliation[WHU]{organization={School of Computer Science, Wuhan University},
  city={Wuhan}, country={China}}

  \affiliation[UMU]{organization={Department of Computer Science, ADSLab, Umea University},
  city={Umea}, country={Sweden}}

\tnotetext[1]{Preprint submitted to the \textit{Journal of Systems and Software} (JSS).}

\begin{document}

\begin{abstract}
Quantum systems are emerging as a transformative technology, harnessing the principles of quantum mechanics through programmable quantum bits (QuBits) to deliver, in certain cases, superior computational capabilities termed as quantum supremacy in computing. In recent years, academic research, industry led projects (e.g., Amazon Braket, IBM Quantum) as well as international consortiums such as the `Quantum Flagship' are aiming to develop solutions for commercially viable quantum computing (QC). In this context, Quantum Computing as a Service (QCaaS) represents a service-oriented solution that provides access to quantum computing resources and platforms as utility-based services, enabling individuals and organisations to use QC capabilities without owning or managing quantum hardware and software. This research employs a process-centric and architecture-driven approach to offer a software engineering perspective on enabling QCaaS -- a.k.a quantum service-orientation. To conduct this research, we followed a two-phase research method comprising (i) a systematic mapping study and (ii) an architecture-based development, first to identify the phases of the quantum service development lifecycle and subsequently to integrate these phases into a reference architecture that supports QCaaS. The systematic mapping study (SMS) process retrieved a collection of potentially relevant research literature and based on a multi-step selection and qualitative assessment, we selected 41 peer-reviewed studies as published evidence to answer three research questions (RQs). The RQs investigate (a) \textit{demographic details} in terms of frequency, types, and trends of research, (b) phases of \textit{quantum service development lifecycle} to derive a reference architecture for conception, modeling, assembly, and deployment of services, and (c) \textit{emerging research trends} that can shape future practices on software engineering for QCaaS. The results identify a 4-phased development lifecycle along with quantum significant requirements (QSRs), various modeling notations, catalogue of patterns, programming languages, and deployment platforms that can be integrated in a layered reference architecture to engineer QCaaS. Emerging and future research efforts focus on topics such as classifying the QSRs, enabling model-driven or low-code quantum service engineering, and integrating large language models in the quantum service development process. In the broader perspective of quantum software engineering (QSE), the results of this SMS specifically focus on quantum service-orientation by providing empirically grounded findings, highlighting patterns for knowledge reuse, and offering a reference architecture as a blueprint to engineer QCaaS.
\end{abstract}


\maketitle

\section{Introduction}
\label{sec:introduction}

Quantum computing (QC) aims to revolutionise the existing computing paradigm by operating on the principles of quantum mechanics, unlike conventional systems that rely on digital circuits and modular applications \cite{R1_ali2022software}. QCs utilise quantum bits (QuBits), which can be programmed to manipulate quantum gates (QuGates) to perform computational tasks that, in certain cases, surpass the capabilities of classical digital computers -- a phenomenon known as quantum supremacy in computing \cite{QuantSupermecy}. QC systems are continuously evolving, and although still in their early stages due to hardware limitations or lack of software ecosystem(s), such systems have started to computationally outperform classical computers in certain application areas, including quantum information processing, bio-inspired computing, and quantum simulations \cite{QuantSupermecy}. Academic research \cite{R1_ali2022software}\cite{R3_egger2020quantum} and industrial efforts led by leading QC vendors such as IBM, Google, and Microsoft \cite{R4_dyakonov2019will} \cite{GoogleWillow} are competing to secure strategic advantages in quantum technologies, often described as the race to the quantum economy. According to the World Economic Forum's report titled \textit{State of Quantum Computing: Building a Quantum Economy}, public and private investments in quantum computing reached \$35.5 billion by 2022 \cite{R5_WorldEconomicForum}, rising to \$55 billion approx. globally in 2024 \cite{Quantum2024}. Despite the strategic potential of quantum supremacy, programming, operating, and maintaining quantum computers presents a complex and fundamentally different engineering challenge from classical computing and software engineering \cite{R1_ali2022software}. QC specific challenges include but are not limited to quantum domain engineering, circuit modeling, algorithm design, and quantum-classic hybrid system deployment. These challenges stem from a lack of systematic processes, insufficient tool support, and a shortage of expertise in quantum software and system development. To address such challenges, state-funded projects and global initiatives such as the Quantum Flagship \cite{R14_riedel2019europe} and National Quantum Initiative \cite{R22_raymer2019us} are investing in developing hardware components, software ecosystems, networking technologies, and skilled human resources for the anticipated quantum leap in computing \cite{R13_monroe2018quantum}.

\vspace{0.3em}

\begin{figure}[]
 \centering
 \includegraphics[scale=0.75]{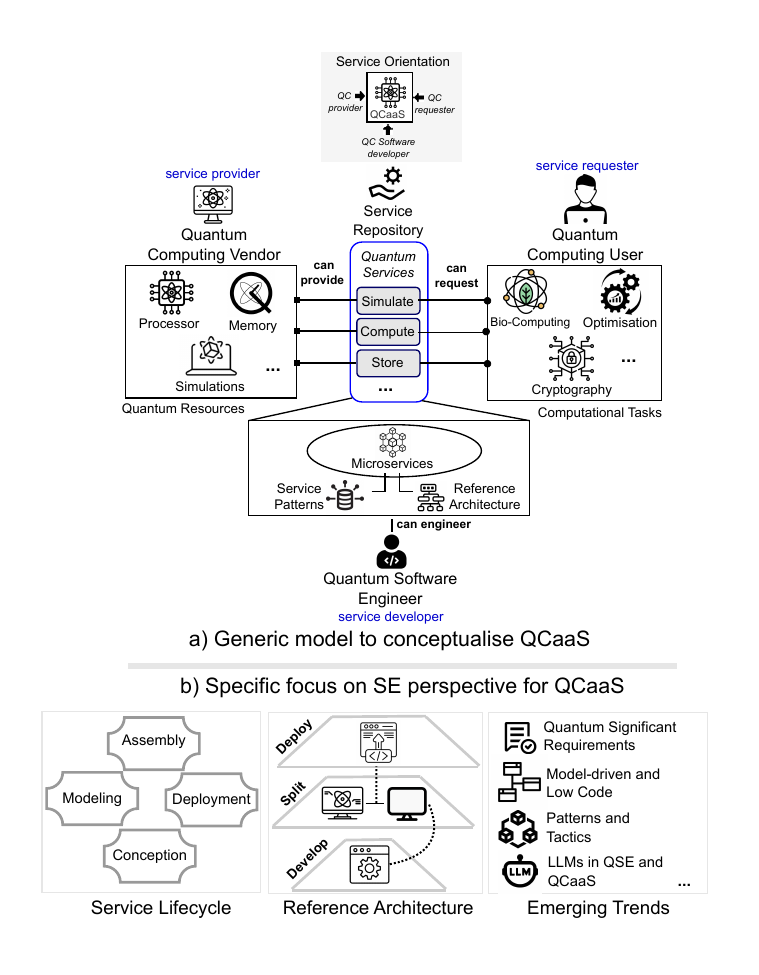} 
 	\caption{Study Focus: A Generic SE view of QCaaS Model}
	\label{Fig-2:QCaaS}
\end{figure}

\textit{Research context and motivation:} Service-oriented systems support the utility computing model, enabling individuals and organisations to access multiple computing services on a pay-per-use basis without owning or maintaining them \cite{R6_wei2010service}. Since their inception and early adoption, pay-per-use, service-driven systems have grown from data storage, video streaming, and entertainment to resource-sharing applications, forming a multi-billion-dollar segment of the service economy \cite{R7_bouguettaya2017service}\cite{R8_Statista}. The `as-a-Service (aaS)' model has driven the widespread adoption of platforms that provide end users with various computing services -- including storage, computation, infrastructure, platforms, and software -- all delivered through on-demand access \cite{R7_bouguettaya2017service}. The aaS model also enables developers and users to utilise QC platforms (processors, memory, simulators) provided by vendors such as Amazon, Google, and IBM \cite{R9_moguel2022quantum}. Quantum Computing as a Service (QCaaS), conceptualised in Figure \ref{Fig-2:QCaaS}, represents a quantum-specific extension of the aaS model. It follows the service-oriented philosophy of pay-per-use, allowing access to quantum resources without the need to own, program, or maintain quantum computers \cite{R10_garcia2021quantum}. For QC vendors, pay-per-shot, where a shot is a single execution of a quantum task on a quantum processing unit (QPU), serves as a revenue-generating business model. In simple terms, we define QCaaS as a `quantum-specific variation of the aaS model that offers quantum hardware and software resources such as processor, memory, algorithms, and simulators etc. that can be utilised by users (via client nodes) by connecting to a remotely deployed quantum computer (i.e., server)'. The QCaaS model can alleviate the need to own or maintain QCs, allowing software and service developers to build applications on QC platforms by leveraging existing knowledge and best practices \cite{R11_leymann2020quantum}. Quantum software services allow developers to encapsulate data and computations in loosely coupled, fine-grained modules to execute tasks on QC platforms \cite{R10_garcia2021quantum}. As shown in Figure \ref{Fig-2:QCaaS}, service developers can leverage reusable knowledge and best practices from classical service engineering such as microservices frameworks and service design patterns to implement quantum services efficiently and effectively. The developed quantum services can enable quantum users, i.e., service requesters, to utilise QC resources offered by quantum vendors, i.e., service providers via pay-per-shot QCaaS model.Without a systematic and structured approach, and in the absence of sufficient expertise in QC service development, there is are risks of inefficient QuBit utilisation, suboptimal quantum–classical task allocation, and error-prone circuit design, all of which can ultimately hinder the success and sustainability of QCaaS. \cite{ServiceDeploy2024}.

\textit{Software Engineering for QCaaS:} Quantum software engineering, as a quantum-specific genre of classical software engineering, leverages established concepts such as architecture models as system blueprints and patterns as reusable knowledge to support software service developers in engineering solutions for QCaaS \cite{SECQSE}. By applying SE best practices, developers can, for instance, utilise the quantum-classic split pattern to design hybrid applications, map architectural components to quantum circuits, and enable model-driven quantum code generation \cite{X2_QSA} \cite{QSEPractationers}. In this research, we have used an SE-focused two-phase research method that relies on a systematic mapping study \cite{SMSEmpirical} to investigate existing literature and follow the architecture-based development \cite{ABD} for an incremental design, implementation, and deployment of quantum software services. Systematic mapping studies (SMS) use an evidence-based software engineering approach to systematically identify, analyse, synthesise, and document research trends on a given topic \cite{R16_petersen2008systematic}. The academic community leverages QC platforms to advance research and develop prototypes \cite{R9_moguel2022quantum}, while quantum computing vendors use them to generate revenue and test their evolving quantum systems \cite{R10_garcia2021quantum}. By synergising research and development on quantum computing with service-orientation, researchers and developers can apply established design patterns, architectural styles, and modeling techniques to deliver QC as a service \cite{R11_leymann2020quantum}. In the context of QSE, this research synthesises the available evidence reflected via published research to provide a software engineering perspective on architecting, implementing, and operationalising QCaaS. We conducted this study to answer three research questions that investigate (i) \textit{demographics:} trends, types, frequency and progression of research (ii) \textit{quantum service lifecycle:} that can be integrated in a layered reference architecture for conceptualising, modeling, assembling, and deploying services, and  (ii) \textit{emerging research:} to highlight topics for potentially futuristic research on QCaaS. 
The results indicate that in quantum service-orientation, QSRs, both functional aspects and quality attributes, guide the four-phase service development lifecycle. Software modeling languages (e.g., UML, circuit diagrams) and patterns (e.g., API gateway, quantum-classic split) can be applied within the proposed layered architecture of QCaaS via quantum-classical hybrid computing. Emerging trends include QSR classification, model-driven development, empirically derived patterns, and defined human roles in QCaaS. The main contributions and implications of this research are to: 


\begin{itemize}
    \item analyse and synthesise the published research as available evidence on principles and practices of software engineering that can be applied to QCaaS that enables QSE for services computing.  
    \item identify and document service lifecycle phases into a reference architecture, acting as a service development blueprint, to support classical and quantum-specific service-orientation techniques for QCaaS.
    \item highlight prominent trends - identifying existing gaps that reflect the dimensions of future research to address emerging challenges and support the development of next-generation QCaaS solutions.
\end{itemize}

This research represents one of the pioneering efforts to rely on evidence-based software engineering guidelines to investigate QCaaS and its implications on QSE. The results can offer academic researchers a mapped overview of existing solutions, open challenges and directions for future QCaaS research \cite{R1_ali2022software}\cite{R14_riedel2019europe}. Practitioners can also benefit by leveraging insights on patterns (reusability), notations (solution representation), and implementation (prototyping) to design and architect QCaaS solutions \cite{R9_moguel2022quantum}.

\vspace{0.5em}
Technical context of QC and QSE is presented in Section \ref{sec:ResearchContext}. Research method is discussed in Section \ref{ResearchMethod}. Results of the study to answer the RQs are presented from Section \ref{sec:RQ1Demography} - \ref{sec:RQ3Emerging}. Review of the existing research is presented in Section \ref{sec:Related}. Implications and validity threats are discussed in Section \ref{sec:ImplicationsThreats}. Conclusions and future work are presented in Section \ref{sec:conclusions}.

\section{Research Context}
\label{sec:ResearchContext}
\begin{figure*}[]
 \centering
 \includegraphics[scale=0.66]{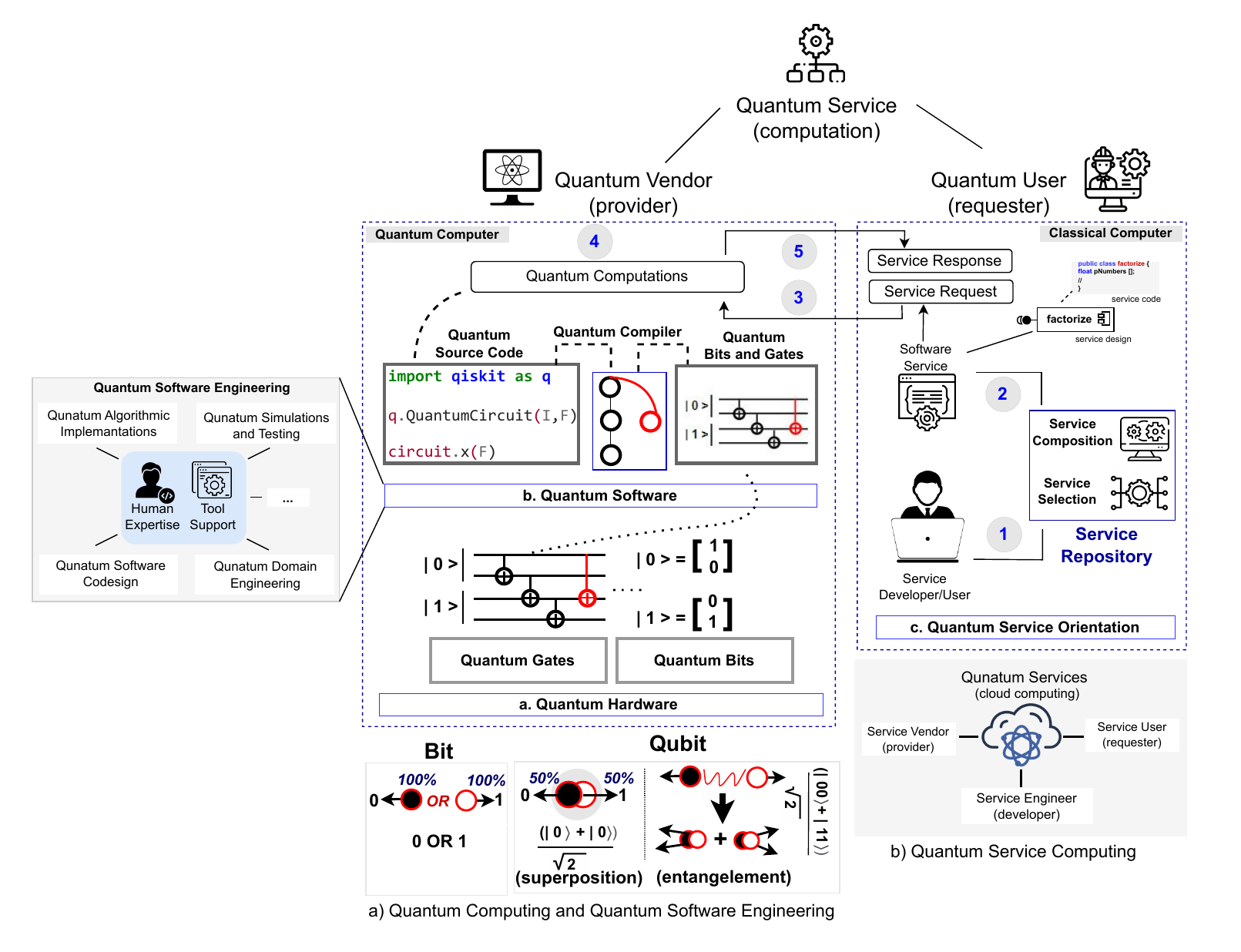} 
 	\caption{An Illustrative Overview of Quantum Computing and Service-Orientation}
	\label{Fig-1:Context}
\end{figure*}

This section contextualises the core concepts used in our study and elaborates them based on illustrations in Figure \ref{Fig-1:Context}. First, we overview the fundamental elements of quantum computing and the role of quantum software engineering to operationalise QC systems in Section \ref{sec:context}, as shown in Figure \ref{Fig-1:Context} a). Second, we explain service-orientation from quantum software engineering perspective in Section \ref{sec:services} via Figure \ref{Fig-1:Context} b). The concepts and terminology introduced in this section are used consistently throughout the paper. They provide a foundational overview of the topic, supported by key references for readers seeking further detail.

\subsection{Quantum Computing and Quantum Software Engineering}
\label{sec:context}
\subsubsection{Quantum Computing Systems}
We explain the QC systems from both hardware and software perspectives, as illustrated in \ref{Fig-1:Context} a). This figure, adapted from \cite{X2_QSA}, has been extended to emphasise software engineering and architecture within the context of quantum service-orientation. At the core of quantum computations is the concept of QuBits, the fundamental units of quantum information, which operate by manipulating QuGates \cite{QuantSupermecy} \cite{R13_monroe2018quantum}. In classical digital systems, binary digits (Bits) are represented as [1, 0], where 1 denotes the ON state and 0 denotes the OFF state, controlling binary gates in digital circuits. In contrast, a QuBit represents a two-state quantum system, expressed as $|0\rangle$ and $|1\rangle$. The state of a single QuBit can be expressed as  $|0\rangle = \begin{bmatrix} 1 \\ 0 \end{bmatrix}$ and $|1\rangle = \begin{bmatrix} 0 \\ 1 \end{bmatrix}$ and quantum superposition allows a QuBit to attain a liner combination of both states, expressed as: 

\begin{equation}\label{EQ-1}
 |0\rangle  =  \left[ \begin{array}{c} 1 \\ 0 \end{array} \right] ~~~~~ + ~~~~~  |1\rangle  =  \left[ \begin{array}{c} 0 \\ 1 \end{array} \right]   
\end{equation}

As shown in Figure \ref{Fig-1:Context} a), a Bit can hold a value of either OFF:0 or ON:1 with 100\% probability. In contrast, a QuBit can exist in the state $|0\rangle$, $|1\rangle$, or in a superposition, such as 50\% $|0\rangle$ and 50\% $|1\rangle$. Additionally, two QuBits can become entangled, meaning that measuring one QuBit instantaneously reveals the state of the other QuBit. For more detailed discussions on QuBits and QuGates in QC systems, please refer to \cite{R1_ali2022software} and \cite{R9_moguel2022quantum}. To harness quantum computing resources like processors and memory, control software is required to program QuBits and manage QuGates within a QC system. To support systems and services for quantum computing, there is a need for systematic processes that incorporates tools, models, patterns, frameworks, and human expertise as engineering practices to develop quantum software \cite{R19_ahmadtowards} \cite{R21_perez2020towards}. 

\subsubsection{Quantum Software Engineering}

 QSE has emerged as a quantum-specific extension of classical software engineering \cite{QuantumMeetsSE}. QSE integrates the principles of quantum mechanics with software engineering practices to design, develop, validate, and evolve software systems, services, and applications for the quantum era computing \cite{R15_de2022software}. It can enable software designers and developers with structured processes, architectural models for design visualisation, and reusable patterns, enabling them to engineer software that can run effectively on quantum computers. Quantum software systems depend on quantum source code compilers, which allow algorithm designers and programmers to develop and execute software on QCs. For instance, developers can use quantum programming languages like Q\# by Microsoft or Qiskit by IBM along with a quantum compiler such as Qiskit transpiler to perform programmable quantum computations \cite{R4_dyakonov2019will} \cite{R15_de2022software}. Quantum software systems, capable of managing and controlling quantum hardware, have applications in areas including quantum cryptography, bio-inspired computing, and quantum information processing \cite{R1_ali2022software}. However, the limited availability of quantum hardware, the shortage of skilled quantum software professionals, and the high costs of owning or maintaining quantum computers pose significant challenges to the commercial viability of quantum computing \cite{QuantSupermecy} \cite{R3_egger2020quantum}. In response, vendors offering QCaaS platforms adopt a service-oriented approach, providing quantum resources to customers as a utility computing model \cite{R6_wei2010service} \cite{R9_moguel2022quantum}.

\subsection{Service-Orientation for Quantum Computing} \label{sec:services}

\subsubsection{Services Computing} It adopts a service-oriented architecture (SOA) approach, enabling users to discover and access a variety of software services that encapsulate computing resources and applications provided by vendors \cite{R7_bouguettaya2017service}. Fundamental to the concept of service computing is software service that represents a self-contained, reusable, and platform-independent software module that encapsulates specific functionality or business logic, accessible over a network through standardised protocols, typically as part of SOA \cite{R20_keen2006patterns}. More specific types of services can be web services or microservices that enable Software as a Service (SaaS) model in which software applications are hosted and managed by a provider, accessed by users over the network via a web client, and offered on a subscription basis thus eliminating the need for local installation and maintenance \cite{R10_garcia2021quantum}.
Figure \ref{Fig-1:Context} b) illustrates a SOA-based quantum service model, where a QC user (service requester) accesses quantum resources provided by vendors (service providers) through quantum services.

\subsubsection{Quantum Services Computing}
Despite their capabilities and advantages, current QC systems struggle to execute quantum algorithms that involve large and diverse datasets \cite{QuantSupermecy} \cite{R13_monroe2018quantum}. As shown in Figure Figure \ref{Fig-1:Context} a), handling large volumes of data requires more QuBits and complex QuGates, resulting in deeper quantum circuits and higher error rates, a challenge commonly referred to as noisy intermediate-scale quantum (NISQ) \cite{R4_dyakonov2019will}. To mitigate issues like NISQ, the quantum-classic split pattern divides quantum applications into classical modules (for pre- and post-processing) and quantum modules (for quantum computation), creating qunatum-classic hybrid applications \cite{R12_valencia2022quantum}. A notable example of such hybrid computing is Shor’s algorithm \cite{Shor2024}, which uses quantum computation to find prime factors of integers, with important applications in computer security and cryptography. QC vendors such as Amazon, IBM, and Google now provide access to their QC systems and infrastructure via quantum services \cite{R9_moguel2022quantum}. For instance, Amazon Braket allows users to build, execute, and simulate quantum software using a pay-per-shot model. Current research and development continues to focus on delivering algorithms, hardware, simulations, and mathematical problem-solving capabilities as services on quantum platforms \cite{R10_garcia2021quantum}.

\textbf{Classical vs. quantum microservices:} In the service computing model, microservices allow software designers and developers to structure applications as loosely coupled, deployable modules that handle computation and data storage \cite{Microservices2018}. This approach enables enterprises to adopt service-oriented models by migrating or modernising monolithic applications into collections of smaller, independent services \cite{R20_keen2006patterns}. Industry-leading service providers like Amazon, Google, and Netflix illustrate how microservices architecture supports core business functions such as video streaming or online shopping, while providing scalable solutions that serve millions of users worldwide. \textit{Quantum microservicing} is a term that refers to microservices that are deployed and executed on quantum
computing platforms. For instance, the solution in \cite{QFAAS2024} uses Qiskit, an open-source quantum software development kit, to implement and deliver quantum computations as a set of microservices that run on the IBM Quantum platform.  From an implementation perspective, focused on writing and executing source code, classical and quantum microservices are similar, as both encapsulate algorithms or code modules that can be executed on their respective platforms \cite{R9_moguel2022quantum}. However, from an operational and deployment perspective, which involves managing service execution on quantum platforms, quantum microservices must meet specific software requirements, encompassing both the implemented functionality and the desired quality attributes of the service.

We conclude that, in order to harness QC as utility computing, there is a need to tailor existing principles and methods of service-orientation or develop new architectures, frameworks, and empirically grounded processes to synergise QC and SOA in the context QCaaS. QSE can provide the conceptual foundations, architectural blueprints, and reusable patterns to manage the complexities of QCaaS. By adopting SE principles such as the quantum-classical split, mapping of architectural components to quantum circuits, and model-driven code generation, developers can design and implement hybrid applications that synergise classical and quantum computations. From a utility computing perspective, quantum service-orientation has the potential to bridge the ``quantum divide” -- a challenge highlighted at the World Economic Forum 2023 -- between entities that possess quantum computing systems and infrastructure and those that do not \cite{R24_WorldEconomicForum}. QCaaS also lays the groundwork for quantum computing clouds (QC cloud or simply QCloud), where technology leaders like IBM provide quantum resources to users through a cloud-based model \cite{QuantumCloud}.

\begin{tcolorbox} [sharp corners, boxrule=0.1mm,]
\faEdit \scriptsize{~\textsf{\textbf{Quantum software engineering} as a discipline refers to the systematic application of software engineering principles to quantum computations by adapting engineering practices to hybrid classical–quantum environments and to the constraints of emerging quantum hardware. For example, the Qiskit SDK ecosystem demonstrates QSE in practice, as it provides structured frameworks for designing quantum circuits, integrating them with classical workflows, and testing their execution across simulators and real quantum processors.}}

\scriptsize{~\textsf{\textbf{Quantum service computing} refers to the application of service-oriented principles to the design, composition, and execution of quantum computing functionalities, where quantum resources and capabilities are encapsulated as interoperable services. For example, Amazon Braket provides qunatum service computing by exposing quantum resources (e.g., circuit simulators and quantum hardware backends via multiple vendors) as services, enabling developers to compose QC tasks.}}

\scriptsize{~\textsf{\textbf{Quantum software service} refers to the abstraction and delivery of quantum software capabilities, such as algorithms, compilers, or simulators etc., as modular services that can be executed on quantum platforms (simulators or physical processors) through service-oriented architectures. For example, in the IBM Quantum Experience, algorithms such as Grover’s search are exposed as cloud-based services accessible via APIs like Qiskit, allowing developers to seamlessly invoke and run them on both simulators and actual quantum hardware without managing QC complexities.}}

\end{tcolorbox}

\section{Research Method}
\label{ResearchMethod}

\begin{figure*}[]
 \centering
 \includegraphics[scale=0.95]{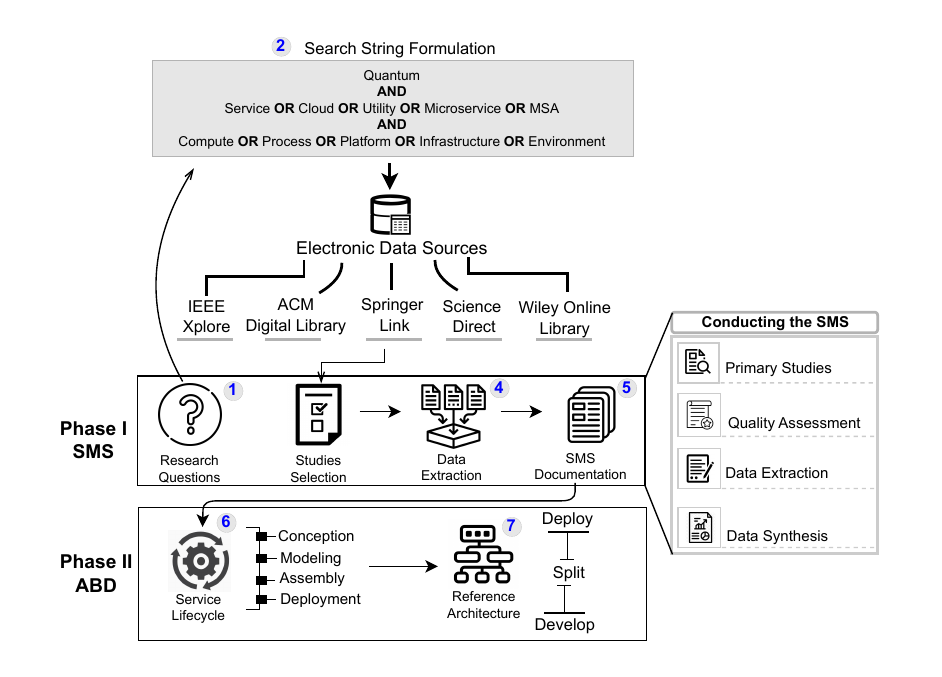} 
 	\caption{An Overview of the Research Method}
	\label{Fig-3:Method}
\end{figure*}

To conduct this study, we used a two-phase research method including (a) systematic mapping and (b) architecture-based development, detailed below and illustrated in Figure \ref{Fig-3:Method}.

\begin{itemize}
    \item \textit{Systematic Mapping Study (SMS):} First, we followed the guidelines and steps of a systematic mapping study that relies on evidence-based software engineering approach to identify, analyse, classify, map, and document available evidence on the topic under investigation \cite{SMSEmpirical}. According to the guidelines in \cite{R16_petersen2008systematic}, a systematic mapping study offers `a means of evaluating and interpreting all available research (as existing evidence) relevant to a particular question, topic area, or phenomenon of interest'. To ensure that our research method adheres to the recommendations and guidelines for conducting the mapping study \cite{R16_petersen2008systematic}, we followed a multi-step process that is illustrated in Figure \ref{Fig-3:Method} and detailed in the remainder of this section. It is worth mentioning here that in recent years, there is a growing interest in conducting multivocal literature reviews (MLRs), which incorporate grey literature (often non-published, non-peer reviewed material), especially in fast-evolving fields like quantum computing and quantum software engineering \cite{MLR}. In this study, we have chosen the SMS approach, adhering to the guidelines of conducting systematic reviews \cite{SMSSLR}, to focus solely on peer-reviewed published research as the primary source of evidence on quantum service computing. Although non-peer-reviewed studies and gray literature such as \cite{R11_leymann2020quantum} \cite{X1_QSE} are considered in this paper to provide additional context for the review of related research (Section \ref{sec:Related}) that complements the findings of primary studies (Appendix \ref{sec:appendix}). Non peer-reviewed and non-published studies are treated as supplementary sources and are not included in the list of primary studies for this research.

    \item \textit{Architecture-based Development (ABD):} As part of SE process, ABD is also referred to as architecture-centric software development \cite{ABD1} where architecturally significant requirements \cite{ASRs} are modelled as system blue-print to guide the subsequent phases of ABD, such as the conceptualisation, implementation, and evaluation of software to be engineered. ABD is generally classified into three main activities, namely \textit{architectural analysis}, \textit{architectural synthesis}, and \textit{architectural evaluation} \cite{ABD}. The first activity, architectural analysis, is driven by a scenario or architecture story as a description of the problem or potential use-cases for the architecture. The next activity architectural synthesis consolidates the details in the story to create an architecture model or representation that can act as a point of reference, visualising the structural (de-) composition of the software. Finally, the last activity architectural evaluation evaluates the synthesised architecture against architecturally significant requirements based on scenarios from the architectural story.
\end{itemize}

In the following, first, the steps of the SMS process are detailed from Section \ref{sec:SMSPlan} to Section \ref{sec:SMSDocument} that follow ABD, which is discussed in Section \ref{sec:SMSDocument}.

\subsection{Step 1 - Specifying the Research Questions} \label{sec:SMSPlan}
As the initial step of the research method, we outlined three research questions (RQs) whose answers represent the results of this research. While formulating these RQs, as per the guidelines of systematic mapping studies \cite{R16_petersen2008systematic} and systematic reviews \cite{SLRSLR}, we focused on three key aspects. These include i) analysing the demography of published research, ii) synthesising the collective evidence on quantum service engineering, i.e., service lifecycle activities, and iii) highlighting prominent trends of emerging challenges/solutions. Each of the RQs is elaborated by discussing the core objectives associated with each question.

\begin{tcolorbox} [sharp corners, boxrule=0.1mm,]
\small
\textbf{RQ-1}: What are the demographic details of published research on software engineering for quantum computing as a service?
\end{tcolorbox}
\textit{Objectives} - The demographic details can include diverse information regarding the frequency, types, core contributions of available evidence, i.e., existing published research. Such details help researchers to identify emerging and potentially futuristic trends of publications on quantum service computing. We specifically focus on four key aspects of demographic details that include the \textit{frequency of publications}, \textit{types of publications}, \textit{types of published research} and \textit{application domains} (e.g., use-cases, scenarios) to which research-based solutions have been applied in the context of QCaaS. 

\begin{tcolorbox} [sharp corners, boxrule=0.1mm,]
\small
\textbf{RQ-2}: What software engineering solutions are reported in the literature to enable or enhance quantum computing as a service? 
\end{tcolorbox}

\textit{Objectives} - Software engineering solution refers to SE artifacts that include but are not limited to process, patterns, architectures, frameworks, tools, etc. that support a systematic and incremental approach to architect, implement and/or reengineer software services. A review of SE solutions indicates the extent to which classical or quantum specific software engineering theory and practice can be applied to engineer QCaaS \cite{QuantumCloud}. We have focused on identifying the \textit{SE solutions} and aligning them with four activities of \textbf{service lifecycle} including  \textit{functional and quality aspects} (\textbf{service conception}), \textit{notations and patterns} (\textbf{service modeling}), \textit{implementation and platforms} (\textbf{service assembly}), and \textit{execution platforms} (\textbf{service deployment}) to support QSE in the context of service development for QCaaS. SE solutions and service lifecycle activities provide the foundation to derive a \textit{reference architecture} for QCaaS.

\begin{tcolorbox} [sharp corners, boxrule=0.1mm,]
\small
\textbf{RQ-3}: What are the emerging trends of research on software engineering for quantum computing as a service?
\end{tcolorbox}

\textit{Objectives} - Emerging trends indicate \textit{open challenges} or potentially \textit{futuristic research} that can shape the agenda, provide a roadmap for progressing research and development, and outline new hypotheses to be tested for QCaaS. The answer can help identify the trends that can help pinpoint the prevalent challenges and their solutions to help researchers and practitioners in developing solutions for emerging and next-generation of QCaaS.

\begin{table*}[b!] 
\scriptsize
\caption{Summary of Steps for Screening and Quality Assessment of Studies}
\begin{center}
{\tiny}
\begin{tabular}{|l|}
\hline
\rowcolor[HTML]{F2F2F2} 
\multicolumn{1}{|c|}{\cellcolor[HTML]{F2F2F2}\textbf{Step I -   Study Screening}}                                                                                \\ \hline
S1 - The study does not propose or discuss any solution for QCaaS                                                                                                    \\ \hline
S2 - The study is not published in English.                                                                                                                                                      \\ \hline
S3 - The study is a duplicate, such as a conference paper later extended into a journal article (the updated and/or extended version is considered).                                           \\ \hline
S4 - The study is a secondary study or survey paper (reviewed as supplementary material Table \ref{tab:RelatedWork}).                                                                                                                                            \\ \hline
\rowcolor[HTML]{BDD6EE} 
\begin{tabular}[c]{@{}l@{}}Exclude the study if the answer to any of the criteria in Step I (S1 - S4) results in Yes, otherwise, \\ Include the study for quality assessment in Step II\end{tabular}  \\ \hline
\rowcolor[HTML]{F2F2F2} 

\multicolumn{1}{|c|}{\cellcolor[HTML]{F2F2F2}\textbf{Step II - Quality Assessment}}                                                                    \\ \hline
Q1 - Are the study objectives and contributions clearly stated?  [Yes:1, Partially:0.5, No:0]                                                                                                     \\ \hline
Q2 - Is the research method reported?  [Yes, Partially, No]                                                                                                   \\ \hline
Q3 - Are design and/or implementation details of the solution provided? [Yes, Partially, No]                                                                                      \\ \hline
Q4 - Are details of experiments, evaluation, or demonstration of the solution provided? [Yes, Partially, No]                                                                         \\ \hline
Q5 - Are study limitations and future research directions discussed? [Yes, Partially, No]                                                                                          \\ \hline
\rowcolor[HTML]{BDD6EE} 
\begin{tabular}[c]{@{}l@{}} Exclude any study with a quality assessment score (sum of Q1–Q5) below 2.0; otherwise,  \\ Include the study for review and data extraction in Appendix \ref{sec:appendix} - Table \ref{tab:data-schema}\end{tabular} \\ \hline
\end{tabular}
\end{center}
\label{tab:qualitycriteria}
\end{table*}

\subsection{Step 2 - Search String and Data Sources}

The initial identification of key terms or keywords for literature was carried out by the first three authors through an analysis of the RQs. The goal of identifying the key research terms was to formulate a search string that can be customised as per specific Electronic Data Sources (EDS) for an automated search, as per the guidelines in \cite{SearchString2020}. Ultimately, the authors reached a consensus on the search string in Listing \ref{lst:searchquery}. The key terms were combined using the boolean operators “OR” and “AND” to formulate the search string. This final search string was developed following a pilot search of relevant literature on IEEE Xplore and Google Scholar. The purpose of the pilot search was to uncover existing study titles and various synonyms for software engineering and quantum software engineering in the context of software services. For example, we discovered that using the key term `framework' instead of the key term `process' returned a significant number of irrelevant studies focused on quantum computing frameworks that were related to QC in general with no relevance to QSE or quantum service computing. Moreover, inclusion of key terms such as `qubit' retrieved hardware-focused studies, not software and/or service models, thus adding irrelevant studies (noise). We also dropped the key terms `quantum-classical' and `quantum classic' as both overlapped with the main term `quantum' that sufficiently retrieves most relevant studies and avoids an extensive list of irrelevant studies to be scanned. The primary goal of the final search string was to identify the most relevant literature while minimising the inclusion of irrelevant studies, thereby leading to a manual review of titles, keywords, and abstracts for study selection.











\begin{lstlisting}[caption={Search String to Identify the Primary Studies}, label={lst:searchquery}, basicstyle=\scriptsize\ttfamily]
"Quantum" AND 
("Service" OR "Cloud" OR "Utility" OR "Microservice" OR "MSA") AND 
("Compute" OR "Process" OR "Platform" OR "Infrastructure" OR "Environment")
\end{lstlisting}


When conducting systematic reviews and systematic mapping studies, EDS facilitates automated searches using predefined and often customised search strings to identify relevant literature about the topic under investigation \cite{EDS}.  Following recommendations for implementing a systematic search process and choosing the most appropriate data sources, we selected five EDS for automated search \cite{LitSearch}. These include IEEE Xplore, ACM Digital Library, Springer Link, ScienceDirect, and Wiley Online Library, which are widely recognised as key sources for literature on computing, particularly software engineering and service computing research. While this list is not exhaustive and does not claim to cover all existing literature, prior SLR or SMS-based studies have identified these five electronic sources as sufficiently comprehensive and appropriate for indexing relevant research studies or evidence. 

The search process included execution of the search string in Listing \ref{lst:searchquery} that was customised in the context of each EDS in Figure \ref{Fig-3:Method}. As a specific example, we utilised the advanced search feature which is available on IEEE Xplore including the \textit{`Search Term'} option to execute the search string and find published studies within \textit{`Full Text \& MetaData'} To refine our results and to remove potentially irrelevant studies, we adjusted the search parameter, switching from \textit{`in Full Text \& MetaData`} to \textit{`in Abstract.'} The search conducted between December 10 - January 3, 2025 included all published studies until December 2024, which resulted in a retrieval of titles of 287 research articles representing potential primary studies. As part of this activity, the search results were subsequently filtered by the first two authors based on titles, keywords, and abstracts by following inclusion and exclusion criteria, i.e., study screening in Table \ref{tab:qualitycriteria} that resulted in limiting the relevant studies to 105 studies. Finally, a full-text analysis was conducted, leading to the selection of 55 primary studies for data extraction. Third, fourth and fifth authors were invited to review the obtained search results and the list of selected articles.

Following the automated search via EDS, we employed the snowballing technique to manually examine the reference lists of the 55 selected primary studies, aiming to uncover any additional studies that may have been missed during the initial string-based review process \cite{SnowBall}. The snowballing technique led us to identify two more studies that were screened for inclusion or exclusion. The first and second authors primarily conducted the snowballing technique, while the third and fourth authors were invited to collaboratively verify their findings to avoid potential bias.

\subsection{Step 3 - Study Selection for the SMS}
To select the primary studies for SMS, we followed the recommendations by Kitchenham et al. \cite{SLRSLR} for including or excluding identified studies. Table \ref{tab:qualitycriteria} presents a two-step criteria-driven process for literature inclusion and exclusion via screening and quality assessment. The first step, referred to as screening, enabled us to filter out any irrelevant, redundant, or non-English studies. The second step, namely quality assessment, enabled us to do a qualitative assessment of each included study, eliminating those that did not meet the formulated quality criteria. The established criteria filtered the search results produced by the search string. 


\subsubsection{Searching the Primary Studies} 

Based on automated search and snowballing, a total of 41 studies were ultimately selected for review and analysis to answer the RQs, as in Figure \ref{Fig-3:Method}. The complete list of these primary studies is detailed in Appendix \ref{sec:appendix} - Table \ref{tab:study-list}. 

\subsubsection{Performing Quality Assessment}
The selected studies were evaluated using quality assessment criteria from Table \ref{tab:qualitycriteria} to minimise research bias and assess the significance and completeness of each study. These criteria includes five specific assessment questions (Q1 – Q5), against which each primary study was evaluated. A score of 1 was assigned if a study explicitly addressed the specific Q, 0.5 if the Q was partially addressed, and 0 if there was no evidence of adressing the Q. The total quality score for each study was calculated by summing up the scores assigned for all the QAs. The first author conducted the quality assessment, and the second and third authors independently verified the results for accuracy and consistency. Studies with a cumulative Q score of 1.5 or higher were included in the final list. 

\subsection{Step 4 - Extracting Data from Studies}
A set of data extraction items were defined to extract specific data related to the RQs. These items represent specific data points extracted from each primary study, aligned with the RQs. The template for data extraction is presented in Appendix \ref{sec:appendix} - Table \ref{tab:data-schema}. The first author conducted a pilot extraction on 10 studies (25\% approx.) to evaluate the appropriateness of the data items, which were revised based on feedback from the second and third authors. The first three authors collaboratively performed the formal data extraction. The extracted data was synthesised before documenting the results of the SMS - available via \cite{SMSdataset}.


\subsection{Step 5 - Documenting the Mapping Study} \label{sec:SMSDocument}
The results of the SMS documented as demographic details of published research are in Section \ref{sec:RQ1Demography} (answering RQ-1), software engineering solutions are discussed in Sections \ref{sec:RQ2SESolutions} - \ref{sec:RefArch} (answering RQ-2), and emerging trends are highlighted in Section \ref{sec:RQ3Emerging} (answering RQ-3).

\textit{Supplementary material} for the study is provided as part of the extended details of the documentation. Specifically, the extracted data from primary studies - organised according to the data extraction form presented in Appendix \ref{sec:appendix} provides the basis for answering the RQs and documenting the SMS, and is made available via \cite{SMSdataset}. Moreover, a dedicated GitHub repository and a short video demonstration showcasing the architecture-based development of quantum software are accessible at \cite{QADLGit2025} and \cite{QADLYT2025}, respectively.
\section{Demography of Research on QCaaS (RQ-1)} \label{sec:RQ1Demography}

This section overviews the demographic details of published research to answer RQ-1 based on illustrations in Figure \ref{Fig-3:MDemography}. Specifically, RQ-1 presents demographic details about the years and types of publications (Section \ref{sec:years}), type of research (Section \ref{sec:type}), and application of research solutions to specific domains (Section \ref{sec:domain}) that complement the answers to other RQs. Demographic details can reveal trends such as most recent research, e.g., journal articles [S3][S33]\footnote{The notation \textbf{[SN]}, where \textbf{S} = study and \textbf{N} = an integer between 1 - 41 such as S1 (first study in Appendix \ref{sec:appendix} - Table \ref{tab:study-list}) indicates a unique reference in the list of primary studies selected for the SMS. This notation also helps to distinguish between the selected primary studies and references in the bibliography section of this paper.} published in 2024 propose an innovative solution or evaluate existing research to engineer quantum software for enabling qunatum cloud computing and computational optimisation, further elaborated below. Throughout this paper, while answering the RQs, a summary of the core findings and guidelines representing key takeaway message(s) is provided and denoted with symbol \faEdit.

\subsection{Frequency and Types of Publications}\label{sec:years}
\begin{figure*}[h]
 \centering
 \includegraphics[scale=0.80]{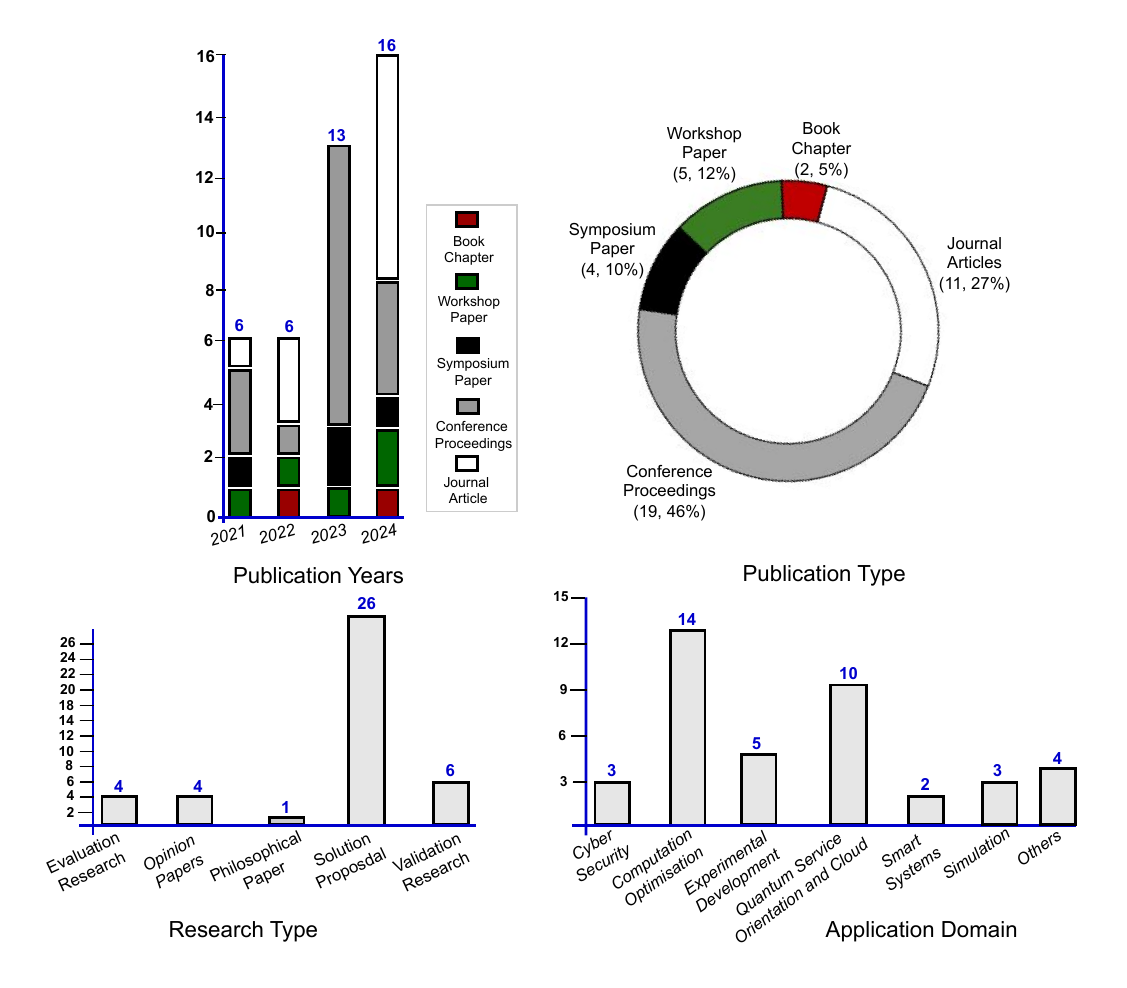} 
 	\caption{Demographic Details of Published Research on QCaaS}
	\label{Fig-3:MDemography}
\end{figure*}

Analysing the frequency and types can highlight the relative growth and trends of publications that indicate a) number of peer-reviewed published studies (i.e., frequency: publications per year) and b) prominent fora of publications (i.e., publication types: journal article, conference proceedings etc.) on topic under investigation. The illustrations in Figure \ref{Fig-3:MDemography} a) and Figure \ref{Fig-3:MDemography} b) visualise the core findings about frequency and types of publications discussed below. Specifically, Figure \ref{Fig-3:MDemography} a) shows four years including 2021, 2022, 2023, 2024, presented along the horizontal-axis, and number of publications in a given year, represented on vertical-axis. It is vital to mention that Figure \ref{Fig-3:MDemography} indicates no published study prior to the year 2021 and the reason for this can be two-fold including (i) our literature search (Section \ref{ResearchMethod}, Figure \ref{Fig-3:Method}) did not identify any relevant studies before 2021, or (ii) any identified study before 2021 did not pass the quality evaluation criteria (Table \ref{tab:qualitycriteria}) for inclusion in the list, i.e., Appendix A - Table \ref{tab:study-list}. The studies [S5] [S2] published as workshop paper and conference proceedings in 2021 represent one of the earliest published research on experimenting with microservices that can be executed on quantum computing platforms. The most recent research, for example as reported in [S10][S18] published in 2024 as journal articles focuses on architecting software services for QC platforms. The bar graph reveals a significant growth of published research in the year 2023 and 2024 (i.e., 29/41 studies, reflecting 71\% approx.) of the selected studies highlighting a progression of the research from experimental solutions [S5, S9] to identifying patterns [S16] and creating reference architecture [S10] that support service engineering for QC software. 

Figure \ref{Fig-3:MDemography} b) indicates five types of publications, also referred to as publication fora classified as Journal articles (11/41 studies, i.e., 27\%), Conference proceedings (19, 46\%), Symposium papers (4, 10\%), Workshop Papers (5, 12\%), and Book Chapters (2, 5\%). Conference proceedings represent the most prominent fora of publications with research focused on publications on quantum function as a service [S3], exploring quantum service-oriantation and cloud computing [S20], quantum microservicing for smart systems [S25], and quantum-classic hybrid computing [S32]. A 2021 report by Scopus also highlights emerging research trends in quantum computing, based on Scopus-indexed documents, with a focus on the demographic details of published research in general area of quantum computing systems, software, and services. The report offers a comprehensive overview, analysing the publication focus, leading institutions, and key contributors in the field of QC \cite{QCTrends2025}.

\begin{tcolorbox} [sharp corners, boxrule=0.1mm,]
\faEdit \scriptsize{~\textsf{\textbf{Publication frequency} There is a significant growth in published research in last 2 years: 2023-2024 (n = 29, 71\%) that propose emerging solutions such as quantum-classic hybrid software, quantum cloud and edge computing as well as leveraging classical service engineering techniques such as quantum SOA, quantum web servicing  that can be customised or reused for quantum software systems.}}

\scriptsize{~\textsf{\textbf{Publication types} Conference proceedings and journal articles are most prominent fora of publications disseminating research via well-established and top-ranked software and service research venues, e.g., JSS, FSE, ICSOC, IEEE Cloud. Some recently organised workshops, e.g., QP4SE conference series, e.g., IEEE QSW and edited book, Quantum Software Engineering represent venues dedicated to publishing theory and practice of quantum software and services.}}
\end{tcolorbox}

\subsection{Types of Research}\label{sec:type}

The type of research, also commonly referred to as research type facet classifies the published studies based on their specific contribution, such as an innovative model, architecture, framework, or prototype as a newly proposed solution [S1, S3, S10] or validation of an existing solution with a new dataset or case studies [S15].  The classification for the type of research is based on si\textit{}x research type facets namely \textit{experience report}, \textit{evaluation research}, \textit{opinion papers}, \textit{philosophical papers}, \textit{solution proposal}, and \textit{validation research} based on classification criteria in \cite{R16_petersen2008systematic}. Our analysis did not find any experience report that documents personal experiences or perspectives of researchers and practitioners on a specific problem and/or project, shown in Figure \ref{Fig-3:MDemography} c). 
Solution proposal is the most prominent type of published research (26 studies, 63\%) that document solutions such as processes [S26, S33], patterns [S16, S18], architectures [S10], frameworks [S3, S1], and prototype [S29] etc. for quantum software services. In contrast, we only identified one philosophical paper [S8] that argues about hybrid nature of quantum services that split computation as classical and quantum components. The paper also sketches a toolchain to synergise classical and quantum computations into a hybrid computing model. Evaluation research (4 studies), opinion papers (4 studies), and validation research (6 studies) are reported in a total of 14 studies, representing a total of 34\% approx. Specifically, evaluation research aims to evaluate an existing solution based on experimentation, such as a case study on quantum service-orientation in smart cities [S11] and evaluating system performance via simulation of distributed quantum computing [S13]. Opinion papers discuss authors’ negative or positive opinions, i.e., pros and cons regarding a specific technique such as orchestration [S18], deployment [S15], and cloud-based [S19] quantum servicing. Finally, the validation research aims to validate the stated hypothesis, objectives, or proposed features of the solution, which have not yet been deployed in a real-world environment. These types of studies focused on validating the integration of quantum-classic microservices [S5], support of quantum algorithm as a service [S6] and enabling the quantum SOA [S7].

\begin{tcolorbox} [sharp corners, boxrule=0.1mm,]
\faEdit \scriptsize{~\textsf{\textbf{Type of research} Analysing the types of published research highlights that solution proposals represent the most prominent type of published research with a total 26 out 41 studies. QCaaS as a recent and continuously evolving phenomena, i.e., subdomain of QSE relies on theory and practices of service-orientation to propose solution for quantum service-orientation. The role of validation and evaluation type research (total of 8 studies) can be vital to empirically assess existing solutions and methods that can be applied to emerging challenges of QCaaS.}}
\end{tcolorbox}

\subsection{Application Domains}\label{sec:domain}
Application domain refers to the usage scenarios, usecases, or system domains to which the proposed research solutions or findings such as quantum service-orientation and cloud computing [S22, S25] or qunatum API gateway can be applied [S1]. For example, the research in [S11, S23] proposes a solution of Quantum as a Service (QaaS) to secure a smart city systems. This refers to quantum service-orientation (i.e., proposed research) applied to smart systems (i.e., application domain). As illustrated in Figure \ref{Fig-3:MDemography}, our analysis classified the existing research into seven research domains including a) cyber security (3/41 studies, 7\% approx.) b) computation optimisation (14, 34\%) c) experimental development (5, 12\%) d) quantum service orientation and cloud computing (10, 25\%) e) smart systems (2, 5\%), f) simulation (3, 7\%), and (g) others (4, 10\%), each exemplified below. Computational optimisation and qunatum service and cloud computing represent the most prominent domains with a total of 24 studies, 59\% of total studies. For example, the studies [S14, S28] focus on optimising the data search problem with quantum search algorithms. Similarly, the studies [S7] focuses on service orientation of quantum resources and enabling the quantum cloud computing [S19, S22]. We found a total of 4 studies that can be classified under others including traffic flow [13] and 5G networking [S40].

\begin{tcolorbox} [sharp corners, boxrule=0.1mm,]
\faEdit \scriptsize{~\textsf{\textbf{Application domain} Research-based solutions on quantum services computing has diverse application domains ranging from optimisation, simulations, smart systems, and quantum clouds. Quantum cloud computing as an emerging domain synergises the cloud computation model with quantum software services to offer QCaaS that can be applied to diverse domains including but not limited to smart systems and cybersecurity solutions.}}
\end{tcolorbox}

\section{Solutions to Engineer QCaaS (RQ-2)}
\label{sec:RQ2SESolutions}

We now answer RQ-2 that aims to investigate the existing SE methods, as published research, to enable or enhance quantum computing as a service. For a fine-grained presentation and systematic interpretation of RQ-2 results, we present the data from reviewed studies based on the following four points and summary of results in Table \ref{tab:DataExtraction} and visualisation of core finding in Figure \ref{Fig:Results}. The results can be interpreted using:

\begin{enumerate}
    \item \textit{Service lifecycle}: Four activities namely conception, modelling, assembly, and deployment are collectively referred to as foundational activities for IBM SOA lifecycle \cite{R20_keen2006patterns}. These \textbf{four generic activities} are used to contextualise and organise \textbf{seven steps} to engineer software services for quantum computing. Specifically, an activity guides about \textit{what needs to be done?} whereas, the step illustrates \textit{how it is to be done?} For example in study [S1], the first activity namely conception provides foundations for system analysts to conceive the idea of developing a service to execute mathematical computations. Functional aspects and quality attributes as two specific steps of conception can guide service developers to consider the required functionality (e.g., quantum service delivery) and quality (e.g., dynamic adaptation) that needs to be ensured in the implemented service. Each of the seven steps to engineer QCaaS are organised under four activities of SOA lifecycle as in Table \ref{tab:DataExtraction}.  

    \item \textit{Extracted data from studies}: The data extracted from primary studies represents the published evidence that is synthesised and summarised in Table \ref{tab:DataExtraction}. Table \ref{tab:DataExtraction} acts as a structured catalogue, detailing each reviewed study that represents published evidence (Study ID) and seven steps of QCaaS under four service lifecycle activities. For example, Table \ref{tab:DataExtraction} can be looked up to quickly identify that study referred to as [S1] focuses on (i) \textbf{conception}  as optimal delivery of quantum service (functional aspect) via a runtime dynamic adaptation and selection of a  specific service (quality attribute) from a pool of available services. To support (ii) \textbf{modelling} of this service UML deployment diagram is used as a notation and API Gateway pattern is applied to manage client request and runtime selection of most suitable service. To (iii) \textbf{assemble} this service, a usecase on process automation and service optimisation is selected and Flask as a Python-based framework is used to implement the service. Finally, for (iv) \textbf{deployment} of the service, Amazon Braket is used as a service delivery platform. 

    \begin{table*}[t]
  \caption{Summary of the Phases and Activities of Quantum Service Lifecycle (identified from the reviewed literature)}
    \label{tab:DataExtraction}
    \centering
    \renewcommand{\arraystretch}{1.5}
    \setlength{\arrayrulewidth}{0.3mm}
    \arrayrulecolor{black}
    \rowcolors{2}{white}{gray!20}
    \resizebox{\textwidth}{!}{
    \begin{tabular}{|c|c|c|c|c|c|c|c|}
        \hline
        \rowcolor{cyan!20} 
        \textbf{Published Evidence} & \multicolumn{2}{c|}{\textbf{Conception}} & \multicolumn{2}{c|}{\textbf{Modeling}} & \multicolumn{2}{c|}{\textbf{Assembly}} & \textbf{Deployment} \\
        \hline
        \rowcolor{cyan!20}
        \textit{Study ID} & \textit{Functional Aspects} & \textit{Quality Attributes} & \textit{Modeling Notation} & \textit{Design Pattern} & \textit{Service Usecase} & \textit{Programming Language} & \textit{Service Platform} \\
        \hline
        S1 & Quantum Service Delivery & Dynamic Adaptation & UML & API Gateway Pattern & Process Automation and Optimisation & Python,Flask & Amazon Braket \\
        \hline
        S2&	Quantum-Classic Hybrid Computing&	Service Integration and Intreoperability& 	Process Models&	Miscelleneous&	Process Automation and Optimisation&	N/A&	N/A\\
        \hline
S3&	Miscelleneous &	Dynamic Adaptation&	Unified Modeling Language&	Quantum-Classic Split&
	Experimental Computation&	Microsoft technology, Q\#&
	IBM Quantum\\
        \hline
S4&	Quantum Experimental Services&	Miscelleneous&
	Unified Modeling Language&	No pattern&	Experimenatal Computation&	Python	&Amazon Braket\\
        \hline
S5&	Quantum Experimental Services&	N/A	& No explicit notation&	Service Composition	Process& Automation and Optimisation&	Python,Flask&	Amazon Braket\\
        \hline
S6&	Quantum Experimental Services&	Service Integration and Intreoperability& 	UML&	API Gateway& Pattern	Algorithmic Exectuion &	Python&	Rigetti\\
        \hline
S7&	Quantum Experimental Services&	Service Integration and Intreoperability& 	Graph Models&	API Gateway& Pattern	Process Automation and Optimisation&	N/A	&Amazon Braket\\
        \hline
S8&	Miscelleneous& 	Service Integration and Intreoperability& 
	Graph Models&	Miscelleneous&	Quantum Simulation&	N/A&	N/A\\
        \hline
S9&	Quantum Experimental Services&	Service Performance
&	No explicit notation&	Service Composition&	N/A	&Python	&Amazon Braket\\
        \hline
S10	&Quantum Service Delivery&	Usability&	UML&Quantum-Classic Split &Cyber Security 	&C\#& Microsoft Azure Quantum \\\hline
S11&	Quantum Service Delivery&	Security 
&Miscelleneous&	Service Composition	&Smart City&	Qiskit, D-Wave libraries	& IBM Quantum \\\hline
S12	&Quantum-Classic Hybrid Computing&	Dynamic Adaptation&	Miscelleneous&	No pattern&	Process Automation and Optimisation	&OpenAPI&	N/A\\\hline
S13	&Quantum Service Delivery&	Service Performance 	&UML	&Miscelleneous	&Process Automation and Optimisation&	N/A	&N/A\\\hline
S14	&Quantum Service Delivery&	Miscelleneous&	UML&	No pattern&	Cyber Security& 	N/A &
IBM Quantum\\\hline
S15	&Quantum Service Delivery	&Usability &	UML&	API Gateway Pattern &	Process Automation and Optimisation&	Python, Flask &
Google Quantum AI\\\hline
S16&	Quantum Cloud and Service Orchestration &	Computational Efficiency&	Process Models&	Quantum-Classic Split
& Experimental Computation &Python, Zato	&
Google Quantum AI\\
        \hline
S17&	Continuous Deployment	&Computational Efficiency
& Process Models	& No pattern &	Quantum search&	Python,Flask&	Amazon Braket\\
        \hline
S18	&Quantum Cloud and Service Orchestration &	Computational Efficiency& Miscelleneous	&API Gateway Pattern &	Quantum search&	Python,Flask & IBM Quantum\\\hline
S19	&Quantum Service Delivery	&Computational Efficiency
&Miscelleneous&	Miscelleneous&	Quantum Simulation& 	Python,Flask&	IBM Quantum\\
        \hline
S20	&Quantum Service Delivery 	&
Reliability& No explicit notation &	Service Facade& 	Quantum Simulation &	Python&	IBM Quantum\\
        \hline
S21	&Continuous Deployment	&
Security& Workflow model &	API Gateway Pattern &	Algorithmic Exectuion &	Qiskit	& N/A\\
        \hline
S22&	Quantum Service Delivery &	Computational Efficiency
& No explicit notation& 	No pattern& 	Process Automation and Optimisation&	N/A	&IBM Quantum\\
        \hline
S23&	Quantum Service Delivery& 	Security&	Workflow model	&API Gateway Pattern &	Process Automation and Optimisation & Python &D-Wave \\
        \hline
S24	&Quantum-Classic Hybrid Computing&	Computational Efficiency
& Miscelleneous&	No pattern	&Experimental Computation &Python, and open sources libraries	&Amazon Braket\\\hline

S25&	Quantum Service Delivery& 	Computational Efficiency&	Miscelleneous&	Service Composition &Experimental Computation
&Qiskit	& Amazon Braket \\
        \hline
S26&	Quantum Service Delivery& 	Computational Efficiency &Miscelleneous&	Service Facade& 	Experimental Computation
&	Python, Flask, D-Wave libraries	&Amazon Braket\\
        \hline
S27&	Quantum Service Delivery &	Computational Efficiency
&Miscelleneous&	No pattern&	Experimental Computation&
N/A	&Amazon Braket\\
        \hline
S28	&Quantum Classic Hybrid Computing&	Computational Efficiency&	Miscelleneous&	N/A	& N/A	&	N/A	&N/A	\\
        \hline
S29	&Quantum service delivery&	Computational Efficiency& Workflow model	&No pattern&	Experimental Computation&
Python&	Amazon Braket\\
        \hline
S30	&Quantum Cloud and Service Orchestration &	Usability &
	Workflow model&	No pattern&	N/A	&N/A	&N/A\\
        \hline
S31	&Quantum Service Delivery&	Usability & 
	Miscelleneous&	Service Facade &	N/A	&Python, D-Wave's Ocean)	&Amazon Braket\\
        \hline
S32&	Quantum Service Delivery &	Useablity&	Miscelleneous&	No pattern	&Process Automation and Optimisation&	Python	&IBM Quantum\\
        \hline
S33&	Quantum Service Delivery &	Reliability& 
	Miscelleneous	&	Service Composition&		Process Automation and Optimisation	&	N/A&		N/A\\
        \hline
S34	&	Quantum Service Delivery 	&	Computational Efficiency&		Workflow model&		No pattern	&	Process Automation and Optimisation	&	Python, Flask, RESTful API	&	
Microsoft Azure Quantum\\
        \hline
S35&		Quantum Service Delivery &		Computational Efficiency &	Miscelleneous	&	No pattern	&	Quantum Simulation&		Python	&	N/A\\
        \hline
S36	&	Quantum Cloud and Service Orchestration &		Computational Efficiency &	No explicit notation&		No pattern	&	Process Automation and Optimisation&		Qiskit	&	IBM Quantum\\
        \hline
S37&	Quantum Service Delivery &	Computational Efficiency
 &	No explicit notation&	No pattern&	Process Automation and Optimisation&	N/A	&N/A\\
        \hline
S38&	Quantum Cloud and Service Orchestration &	Computational Efficiency&	Miscelleneous&	No pattern&	Process Automation and Optimisation&	N/A	&N/A\\
        \hline
S39&	Quantum Service Delivery &	Computational Efficiency&	Miscelleneous&	No pattern	&Process Automation and Optimisation&	N/A&	N/A\\
        \hline
S40&	Miscelleneous &	Security cloud&	Miscelleneous&	No pattern&	Security &	N/A	&N/A\\
        \hline
S41	&Continuous Deployment&	Computational Efficiency	&No explicit notation&	Service Composition	Process& Automation and Optimisation&	N/A&	D-Wave Leap\\
        \hline

    \end{tabular}}
  
\end{table*}

    \item \textit{Summary of core findings}:  Figure \ref{Fig:Results} accumulates a visual representation of the data from Table \ref{tab:DataExtraction} via Figure \ref{Fig:Results} a) to Figure \ref{Fig:Results} g) and highlights key findings and prominent insights from each of the seven steps for developing QCaaS. For example,  Figure \ref{Fig:Results} d) shows that patterns used to model QCaaS can be generally classified into five types, namely API Gateway, Service Composition, Quantum-Classic Split, Service Façade, Miscellaneous patterns such as enterprise service bus and layered architecture. A total of 17 out of 41 studies (i.e., 42\% of reviewed studies) do not explicitly specify any pattern. Service Façade is applied to a total of 8 studies to support service integration and wrapping of classical services to the interfaces of quantum services. Figure \ref{Fig:Pattern} complements the discussion on quantum service patterns by providing a thumbnail view, as a visual reference, for the identified patterns.
    
    \item \textit{Illustrative example}: We also exemplify Shor’s algorithms as an example to architect quantum service-orientation. The example provides a scenario for a step-by-step illustration and elaboration of the four phases and seven activities of service lifecycle to engineer QCaaS, presented in Figure \ref{fig6:Implementation}.  
\end{enumerate}

\subsection{Activity I - Service Conception}
The service conception activity focuses on conceptualising the required functionality and desired quality of the service(s) to be implemented. From service engineering perspective, functional aspect and quality attributes that have a profound impact on the design, implementation, and delivery etc. of software services are collectively referred to as architecturally significant requirements (ASRs) \cite{ASR}. In the context of quantum service-orientation, ASRs also referred to as quantum significant requirements (QSR) that are crucial for the design, development, execution, or integration of QC capabilities within software-intensive systems, applications, and services. QSRs ensure that software can effectively leverage functionality and quality, such as quantum algorithmic implementation, interaction with quantum hardware, or maintain compatibility with hybrid quantum-classical environments. For example, while architecting quantum services \cite{R9_moguel2022quantum} \cite{R10_garcia2021quantum}, QSRs may require a split of a quantum algorithm in terms of classical and quantum computations for executing data inputs, state preparation, and simulations on a combination of classical and quantum processors that represents required functionality. Moreover, such an execution requires efficient utilisation of QuBits and quantum error minimisation as the desired quality. 

\subsubsection{Functional Aspects} 
\begin{figure*}[]
 \centering
 \includegraphics[scale=0.65]{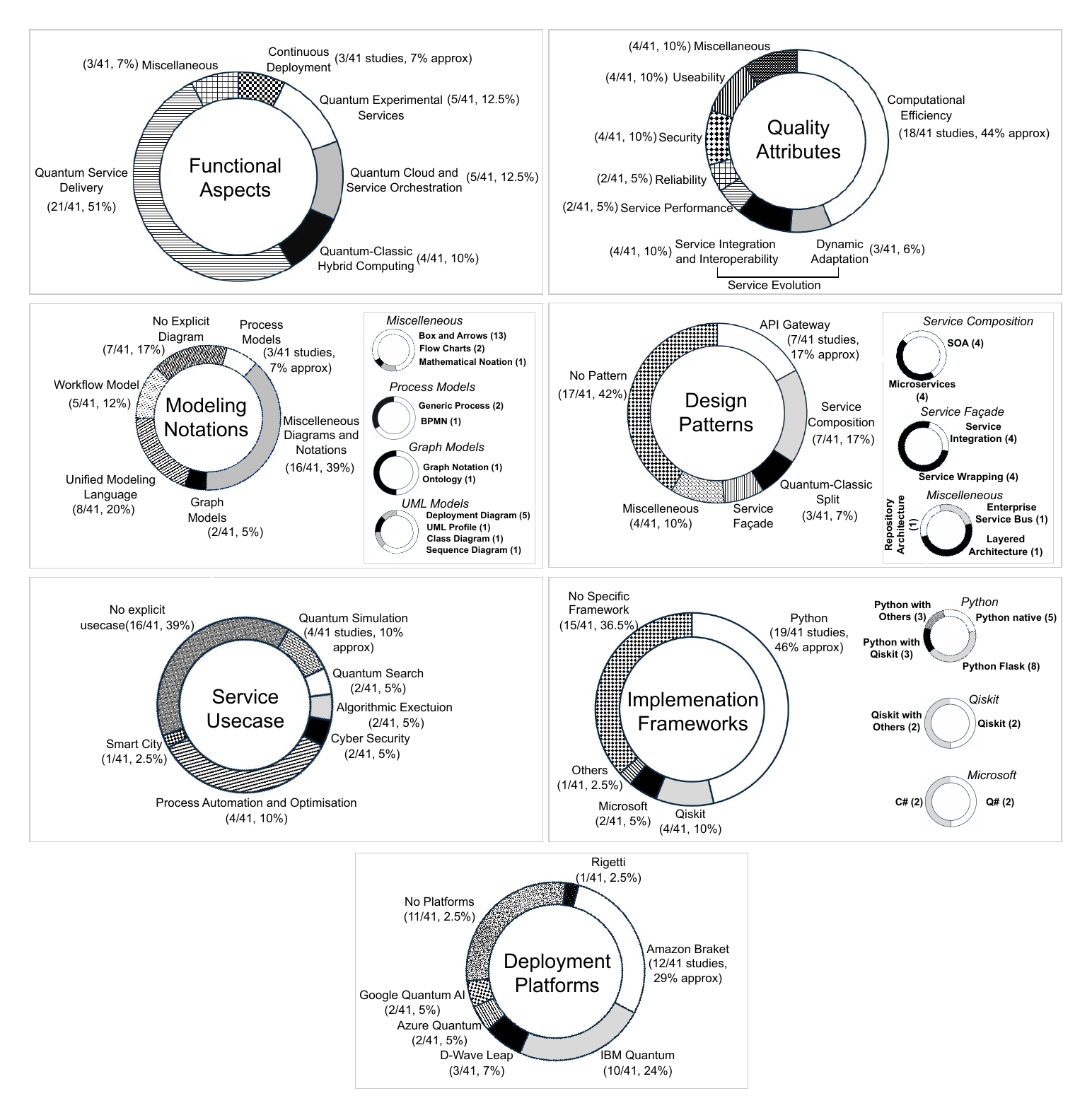} 
 	\caption{An Illustrative Overview of the Results of RQ-2}
	\label{Fig:Results}
\end{figure*}

Based on our review, we identified and classified functional aspects into five (5) specific types while quality attributes are classified into seven (7) specific types as highlighted in Figure \ref{Fig:Results} a). These five types include \textit{Continuous Deployment} (identified in a total of 3 out of 41 (3/41) reviewed studies that represent 7\% approx. of reviewed literature), \textit{Quantum Experimental Services} (5/41, 12.5\%), \textit{Quantum Cloud and Service Orchestration (5/41, 12.5\%)}, \textit{Quantum-Classic Hybrid Computing} (4/41, 10\%), \textit{Quantum Service Delivery} (2/41, 51\%), and \textit{Miscelleneous} (3/41, 7\%). In the context of quantum service patterns, some stuides such as [S10][S16] refer to the quantum-classic hybrid computing as quantum-classic split pattern. For the sake of clarity, the quantum-classic hybrid computing is a general reference to the computing model, whereas quantum-classic split is a specific pattern to implement the hybrid computing model. In the following, we briefly introduce each specific type of functional aspects and quality attributes and exemplify one of them as a representative example for elaboration. For example, considering ‘quantum cloud and service orchestration’ type, the studies [S32][S38] focus on orchestrating software services that can manipulate QC hardware that is deployed on a remote server and made available via the cloud computing model. Specifically, [S38] allows QC vendors to offer quantum resources (e.g., QuBits) that can be utilised via on-demand plans and pay-per-use model, as illustrated in Figure \ref{Fig-1:Context}. As part of service conception, we summarise the functional aspects as:

\begin{itemize}
    \item \textit{Continuous deployment} enabled via quantum microservicing and is identified in a total of 3 out of 41 studies (representing 7\% of all reviewed studies).  By following continuous deployment, the functionality of a service is developed, tested, and deployed for a seamless delivery of software services that can be executed on QC platforms. 
    \item \textit{Quantum experimental} services support proof-of-the-concepts or experimental delivery of software, including but not limited to enabling number crunching, experimental optimisation, or algorithmic execution on a QC platform reported in a total of 3 studies.  Experimental service development offers only a limited functionality, merely capable of checking the status of quantum circuits or generation of random numbers using quantum hardware, referred to as number crunching with QCs. 
    \item \textit{Quantum cloud and service orchestration} enables end-users to invoke software services using cloud computing model identified in a total of 5 studies. Quantum cloud computing is gaining significant attention from leading QC vendors such as IBM cloud (IBM quantum experience) \cite{IBMQCloud} and Google Cloud (Google quantum AI) \cite{GoogleQCloud} with a push for experimentation with quantum algorithms via cloud-based quantum processors. 
    
    \item \textit{Quantum-Classic hybrid computing} focuses on a hybrid computation model that splits a software into two parts namely classical software and quantum software, presented in a total of 4 studies. The hybrid computing model ensures that quantum processing is utilised only where it offers an advantage (so-called quantum supremacy \cite{QuantSupermecy}), while classical computer can handle the routine (classic computation) operations efficiently. In the context of hybrid computing model, we further elaborate on the quantum-classic split pattern based on illustration in Figure \ref{Fig:Pattern}.

    \item \textit{Quantum service delivery} focuses how software is provided, maintained, and accessed by users, whether through SaaS, on-premises, or managed services for QCs, as identified in a total of 21 studies. Conceptually similar to the concepts of continuous deployment and services orchestration, quantum service delivery is different because it focuses on how users access and interact with software services rather than the technical mechanisms of deployment. In contrast, Continuous Deployment (CD) is a DevOps practice that can enable automation of the release of software updates without manual intervention, ensuring frequent improvements to service delivery.

    \item \textit{Miscellaneous} represents a multitude of functional aspects that cannot be classified under a specific type and are identified in a total of 3 studies. These include, for example, the studies [S3] demonstrating the efficiency of data search via QC, [S8] discussing QC applications for bio-inspired computing, and [S40] discussing quantum-safe security.  For example, the study [S8] provides a reference model for QC toolchain and its application to address challenges of bio-inspired computing in the context of humanities and natural sciences.   
\end{itemize}

\subsubsection{Quality Attributes} Quality attributes for software services define the non-functional properties that impact system operations -- supported via a number of measurable characteristics -- including but not limited to performance, security, and efficiency. As a specific example, in a resource-constrained QC platform, efficiency as a quality attribute for QCaaS ensures an efficient utilisation of the available QuBits while executing quantum algorithm as a service [S6]. In service engineering context, the concept of Quality of Service (QoS) is virtually synonymous and often used interchangeably with the concept of quality attributes. QoS refers to the overall performance and reliability of a software system as experienced by users, ensuring that it meets predefined expectations and service level agreements (SLAs) \cite{QOS-2025}. For technical distinction, we can conclude that quality attributes contribute to achieving high QoS, whereas the QoS is the actual service performance, offered by the system and/or experienced by the user. We have identified a total of six types of quality attributes including:

\begin{itemize}
    \item \textit{Computational efficiency} refers to achieving optimal throughput and optimising execution time etc. for scalable and high-performance computing, reported in 18 studies. In the context of efficiency for QCs, the term quantum computational supremacy is often used, referring to a computation threshold where a quantum computer is computationally efficient and outperforms the optimal classical supercomputers in solving a problem \cite{QuantSupermecy}. For example, the study [S18] proposes a quantum orchestrator for load-balancing and efficient execution of quantum tasks on platforms like Amazon Braket and IBM Quantum, supporting both quantum and classical workflows.
    
    \item \textit{Service evolution} refers to the maintainability and adaptability of software services, ensuring their continuous improvement and integration. A total of seven studies have explored this aspect. To provide a more fine-grained analysis, we classify service evolution into two main types: (1) \textit{Dynamic adaptation} (i.e., adaptive evolution), which enables services to adjust to changing requirements, and (2) \textit{Service integration} and interoperability (i.e., perfective evolution), which focuses on enhancing compatibility and seamless interaction between services. In the context of quantum software services, service evolution is particularly critical due to the rapid advancements in quantum algorithms, hardware, and hybrid quantum-classic systems [S7]. Adaptive evolution ensures that quantum services remain efficient despite evolving quantum computing paradigms, while perfective evolution facilitates the integration of quantum services with existing classical infrastructures, promoting broader adoption and interoperability [S3].
    
    \item \textit{Service performance} measures a number of aspects such as resource efficiency, response time, energy consumption etc. to analyse and enhance the performance of quantum services. Service performance aspects can be generally classified as execution time, QuBit utilisation, and ensuring efficient quantum-classical integration, which have been identified in a total of 2 studies including [S9][S13]. For example, the study [S13] proposes a conceptual model integrating blockchain and QC for secure, and efficient serverless edge computing where response time and resource efficiency are identified as the key criteria to measure performance of quantum services. 
    
    \item \textit{Reliability} of quantum software ensures consistent execution, fault tolerance, and error mitigation, etc. to enhance dependable quantum-classical interoperability and robust quantum computing applications - identified in a total of 2 studies including [S20][S33]. As a specific example, to ensure reliability in terms of error-free execution, the study [S33] presents QuantoTrace, a cloud-based Error Correction as a Service (ECaaS) that simplifies quantum error correction for NISQ devices. The QuantoTrace solution aims to enhance quantum service reliability and accessibility while supporting future research in QC strategies.
    
    \item \textit{Security} in the context of quantum service-orientation refers to services that enable secure management and execution of systems and software executed on QC platforms - often referred to as security as a service [S14][S23]. Data security solutions offered via quantum services include robust encryption, quantum-safe cryptography, and resilience against quantum attacks, safeguarding data integrity, confidentiality, and secure communication in quantum-classical hybrid computing environments. Security as a critical quality attribute is identified in a total of 4 studies [S9][S11][S14][S23]. The study [S23] proposes a smart city security architecture that can be deployed on various QC platforms, such as D-Wave and IBM Quantum Platform for real-time data classification and addressing security risks for smart city devices.
    
    \item \textit{Usability} of quantum software services focuses on intuitive interfaces, developer-friendly frameworks, and seamless quantum-classical integration, enabling efficient programming, debugging, and accessibility for researchers and practitioners in quantum computing environments, identified in a total of 4 studies [S10][S15][S30][S31]. For example, the studies [S10][S30] provide a conceptual model, focusing on a user-centric view of quantum cloud computing to enhance usability and availability of QC resources as software services.
    
    \item \textit{Miscellaneous}  category includes a total of 4 studies that either lack an explicit presentation or discussion of any specific quality attributes or present one-off solutions that do not fit into any of the seven existing classification, detailed above. For example, the study [S14] explores services as sustainable software capable of leveraging QC and exploring the sustainability of QCs. The study [S5] focuses on quantum-classical microservices but does not specify any particular quality attribute. 
\end{itemize}

\begin{tcolorbox} [sharp corners, boxrule=0.1mm,]
\faEdit \scriptsize{~\textsf{\textbf{Quantum significant requirements} represent the quantum-specific genre of the classical ASRs to ensure that functional aspects such as quantum-classic split and quality attributes such as efficient QuBit utilisation should be integral part of the conception for quantum SOAs.}}

\vspace{0.5 em}

\scriptsize{~\textsf{\textbf{Functional aspects} of quantum services, classified into 5 main types, mainly focus on leveraging the principles of service-orientation to deliver functional quantum services that can be invoked on QC platforms. A systematic identification and classification of key functional aspects can help to determine the features offered by quantum software services as building blocks of quantum cloud computing, a recent trend, that is gaining attention both from academic research and industrial projects to offer computational resources via pay-per-shot computing model.}}

\vspace{0.5 em}

\scriptsize{~\textsf{\textbf{Quality attributes} are generally classified into 7 types, mainly focusing on non-functional aspects of quantum services. In addition to supporting features as functionality, aspects such as quantum noise minimisation (service reliability), QuBit utilisation (service efficiency) are crucial for ensuring both functionality and overall quality in an attempt to support the notion of quantum computation supremacy.}}
\end{tcolorbox}






\subsection{Activity II - Service Modelling}
After conception, the modelling activity relies on \textit{notations} as a means to represent the design and \textit{patterns} as best practices to create service design based on functional aspects and quality attributes. Modeling notations include but are not limited to graphical, descriptive, or mathematical representation of service design. Patterns complement the notations, as artifacts of reuse knowledge, to model an optimal design. As a specific example, study [S3] uses UML class and sequence diagrams as UML-based modeling notations to represent quantum search algorithms. The quantum-classic split as a layered pattern is applied to modularise the search algorithm in terms of implementing the orchestration of classic and quantum computations. 

\subsubsection{Modelling Notations} These are essential for creating, maintaining, and evolving models such as ontologies and graph-based diagrams, which provide visual representations, while architectural description languages offer textual specifications for software-intensive systems \cite{QUML} \cite{QADL25}. Recent trends in software engineering, emphasising model-driven and low-code application development, have shifted developers’ focus from traditional coding to software modeling for implementation \cite{QLowCode}. The low-code approach applies model-driven engineering principles, treating models as first-class entities throughout the software development process. Moreover, with a recent emergence of large language models as software development bots, there is an increasing reliance on high-level descriptive models or graphical representation to generate source code for qunatum algorithms \cite{QLLM}. Exploring software models and modeling notations that support service modeling is crucial for adopting a model-driven approach in QSE. This approach enables quantum code developers to abstract implementation-specific complexities while creating quantum services \cite{QUML}\cite{QLowCode}. For example, study [S3] presents a class diagram as a structural view of the system, modeling attributes and methods of entities such as users, service providers, and authentication in QCaaS. This SMS identifies five main types of notations for modeling services in QCaaS, as illustrated in Figure \ref{Fig:Results}.

\begin{itemize}
    \item \textit{Process models} in service development lifecycle are used to visualise the workflows, activities, and interactions to analyze, design, and optimise services, ensuring an alignment with business goals, user needs, and requirements implemented as software services. Process models as a modeling notation are identified in a total of 3 studies [S2][S16][S17] that use a process-centric view and business process modelling notation to create service design. For example, in order to abstract out complex and implementation-specific details of quantum algorithmic implemenations, the study [S17] aims to facilitate developers to concentrate on application development rather than low-level quantum hardware details. In order to achieve this, the authors of the study utilised the business process modeling notation (BPMN) to design a process model that is used for generating quantum service source code from API specifications and quantum circuits. 
    
    \item \textit{Graph-based models} have been applied in two studies using ontologies and a directed graph where the graph nodes represent a service and edges represent the interaction between the services [S7][S8]. For example, in [S7] the authors use directed graphs to represents quantum software and graph theory is used as a formal specification language of the quantum program.

    \item \textit{Unified Modelling Language (UML)} based models use a variety of notations inclusing the class and component diagrams that depict structural aspects [S3], while sequence and deployment diagrams capture runtime or behavioral views of software services [S6][S13]. UML diagrams and profiles remain the standard in software modeling and are widely adopted by software and service developers, with increasing application in QSE \cite{QUML}. For example, the study [S10] proposes a reference architecture for quantum computing as a service. A proof-of-the-concept design and implementation of qunatum software services is provided via the Shor's algorithm that is modeled using UML sequence diagram (modeling algorithmic behaviour) and UML deployment diagram (showing the qunatum-classic split) based deployment of the service. The role of UML-based modeling is further exemplified in the context of pattern-based modeling of quantum software services - illustrated in Figure \ref{Fig:Pattern}. 
    
    \item \textit{Workflow model} for service design represents the task to be performed (e.g., functional aspects), roles (entities or actors that interact with the tasks), and interactions to guide the structured development of software. In the reviewed studies, often the terms process modeling and workflow modeling have been used interchangeably, generally referring to modeling a flow of service execution [S16][S34]. From a technical perspective, a distinction between the two needs to be maintained. Specifically, a business process model provides a high-level view of organisational activities, focusing on goals, roles, and processes that can be automated via several services. In contrast, a workflow model details the specific sequence of tasks and data flows, often expressed as flow charts, emphasising service execution and automation. For example, the study [S34] uses workflow modeling to represent the execution and interaction of services to design, implement, and deploy qunatum algorithms.

    \item \textit{Miscellaneous} uses a number of different notations, often referred to as non-conventional modelling in the context of software modelling and design. These miscellaneous notations have been used in a total of 16 studies ustilising box and arrow representations of the architecture, flow charts, and mathematical notations to formally express qunatum service model. For example, the study [S19] utilises architectural overview diagram that is represented via box and arrow noation. A box represents a service in execution, whereas the arrow represents service interaction. In comparison, the study [S38] utilises mathematical notations to express QuBits and quantum circuit utilisation during service execution.

    \item \textit{No exploit notation} represents a scenario where no clear or explicit mention of any modelling notation for quantum services. It is vital to mention that a number of studies have used notation that are specifically developed to model qunatum software. These include qunatum circuit diagram [S10][S38] and quantum architecture description language \cite{QADL25}.
\end{itemize}

\subsubsection{Design Patterns}

Patterns\footnote{In software design and architecture context, the terms \textit{pattern} and \textit{style} are often used interchangeably to refer to best practices or frequently used solutions for recurring software modelling challenges. However, for technical clarity, a style refers to a broad structural blueprint that defines a family of systems by their components and relationships (e.g., client-server style), whereas a pattern is a reusable solution to a specific architectural problem, often addressing component interaction or deployment within a given style (e.g., a layered pattern used to implement the client-server style).} capture the collective expertise of software designers as reusable knowledge and best practices for addressing recurring design and architectural challenges. Given the limited professional expertise in QSE, such as quantum engineers, algorithm designers, and software architects, these patterns serve as reusable artifacts, enabling novice developers to follow established best practices during quantum software development \cite{QPattern}. This SMS highlights that the literature on QCaaS reports five patterns, namely the \textit{API Gateway}, \textit{Layered Architecture}, \textit{Classic-Quantum Split}, \textit{Service Fa\c{c}ade}, and \textit{Repository Pattern} as shown in Figure \ref{Fig:Results}. Figure \ref{Fig:Pattern} depicts pattern thumbnails \cite{PatternThumbnail} that provide a concise visual summary of the pattern, typically including its name and structural representation of pattern elements and their interactions. 

\textbf{Example: API Gateway for Amazon Braket} enables clients to access QC resources via a Gateway -- a pattern used by Amazon as a platform agnostic solution to intermediate between QC users and QC vendors via Amazon Braket \cite{APIGateway}. For example, in {Figure \ref{Fig:Pattern} a), the thumbnail suggests a pattern named \textit{Quantum API Gateway} as a layered system comprises of three main elements including a) Quantum Service Clients that can invoke available services on b) Quantum Server via routing and recommendation by c) Quantum API Gateway, detailed in [S1]. Figure \ref{Fig:Pattern} a) illustrates Amazon Braket that is a managed service provided by Amazon Web Services (AWS), using the API Gateway pattern, for accessing QC resources via QCaaS. The pattern comprises three key layers. At the client node, users aim to access QC resources, having Jupyter Notebook and AWS Console, enabling them to design, simulate, and submit quantum programs via the Braket SDK. The gateway service, i.e., Amazon Braket, acts as an intermediary that abstracts hardware complexity and manages execution workflows, including device selection, queuing, and error handling. Finally, the quantum computing resources layer connects users to multiple quantum hardware providers (e.g., Vendor A, Vendor B, etc.). This model provides a platform-independent interface, facilitating seamless experimentation across various quantum technologies. By decoupling development from hardware integration, Amazon Braket supports scalable, vendor-neutral QC for researchers, developers, and organisations exploring quantum applications.

\textbf{Pattern documentation:} One of the most common mechanisms to document and store patterns are pattern templates and pattern catalogues that provide structured management of available patterns \cite{PatternThumbnail}. Our intent is not to provide comprehensive details on pattern documentation and management, therefore, we rely on pattern thumbnails from Figure \ref{Fig:Pattern} to concisely summarise the identified patterns. We briefly introduce each pattern and elaborate on the quantum-classic split pattern, with an example below. 

\begin{figure*}[]
 \centering
 \includegraphics[scale=0.70]{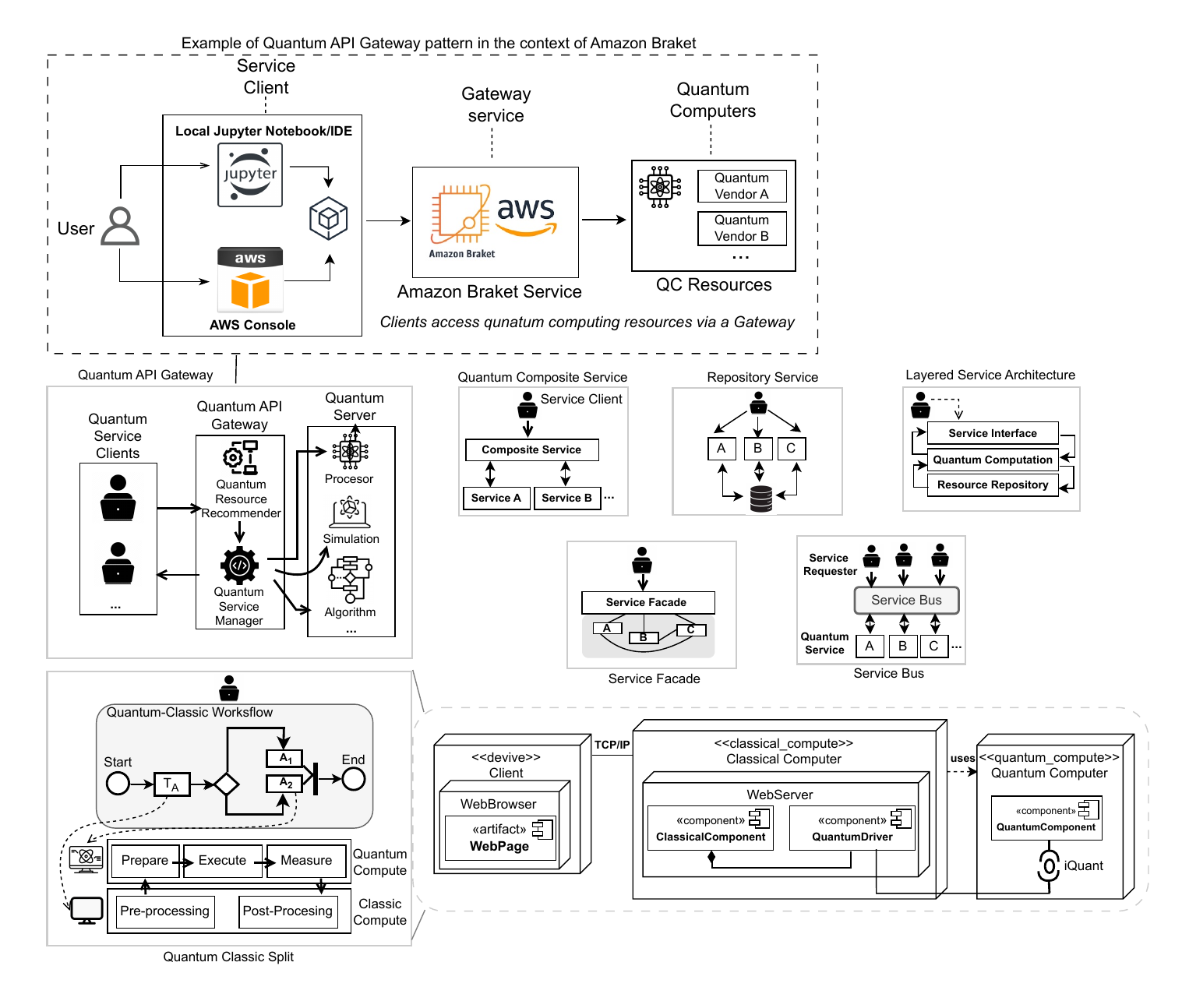} 
 	\caption{Thumbnail View of the Identified Patterns}
	\label{Fig:Pattern}
\end{figure*}

\begin{itemize}

 \item \textit{Quantum-Classic Split} is identified in a total of 3 studies is a quantum version of the Splitter pattern and enables hybrid computation that can be divided between classical and quantum computations [S10][12][s24]. As shown in Figure \ref{Fig:Pattern} c), the classic-quantum split is driven by the workflow, also referred to as the split logic to divide the execution of computations. Figure \ref{Fig:Pattern} shows that the qunatum-classic workflow modeled as a UML activity diagram computes and delegates activities to be executed on classical and quantum computers. For example, the component and connector architectural view in Figure \ref{Fig:Pattern} e) shows that a \textit{ClassicalComponent} and \textit{QuantumDriver} deployed on a classical machine can invoke \textit{QuantumComponent} on a quantum computer. A typical real-world example of the quantum-classic split pattern is Shor's algorithm for prime factorisation. The algorithm divides tasks between classical and quantum computations such that classical processing handles choosing a random integer, and performing the greatest common divisor checks, and postprocessing, whereas the quantum part efficiently determines the period using modular exponentiation and Quantum Fourier Transform (QFT). This hybrid approach leverages quantum speedup for order finding while relying on classical hardware for the remaining computations.

    \item \textit{API Gateway} is identified in a total of 7 studies that focuses on centralising request handling, routing, authentication, and aggregation, acting as a single entry point for microservices communication [S1][S6][S7][S15][S18][S21][S23]. For example, the study [S1] proposes the Quantum API Gateway pattern for adapting traditional API gateways for quantum services by dynamically selecting optimal quantum hardware at runtime. The study also provide a proof-of-concept implementation of the Quantum API Gateway for the Amazon Braket platform. It is vital to mention that Amazon cloud computing infrastructure, i.e., Amazon Braket uses the API Gateway pattern to manage and route requests between users and quantum computing backends \cite{APIGateway}. It acts as an interface where developers submit quantum tasks via APIs, which are then authenticated, queued, and routed to the appropriate quantum processing unit (QPU) or simulator. This pattern abstracts backend complexities, enabling seamless integration of quantum services into classical applications.
    
    \item \textit{Service Composition} pattern combines multiple services into a single, higher-level service, enabling modularity, reusability, and streamlined functionality in distributed systems, identified in a total of 7 studies. For example, the study [S4] utilises a service composition pattern to deploy quantum services on Amazon Braket by wrapping them in classical services, demonstrating hardware variation through integer factorisation across different quantum backends.

    \item \textit{Service Fa\c{c}ade} for quantum software services refers to a classical interface that abstracts and simplifies access to complex quantum functionalities. It acts as a bridge between classical clients and quantum backends, hiding the complexities of quantum computing, such as circuit design, QuBit management, and backend selection, behind a well-organised and easy to implement service-oriented interfaces. Service fa\c{c}ade is identified into a total of 2 studies. For example, the study [S7] utilises the service fa\c{c}ade pattern, which wraps the classical (legacy software services) under the interfaces of quantum microservices that can be executed on Amazon Braket platform.
    
    \item \textit{Miscellaneous} represents a variety of different patterns such as the enterprise service bus [S16], layered architecture [S2][S23], and repository pattern [S7], identified in a total of 4 studies. For example, the study [S2] uses the layered architecture pattern to modularise a qunatum SOA via the process layer, service layer, and infrastructure layer.

\end{itemize}

\begin{tcolorbox} [sharp corners, boxrule=0.1mm,]
\faEdit \scriptsize{~\textsf{\textbf{Quantum service modeling} supports a model-based representation (i.e., creating design or architecture models) of functionality and quality of the conceived services. Modeling phase is supported via:}}

\vspace{0.5 em}

\scriptsize{~\textsf{\textbf{-- Modelling notations} can assist software engineers to shift their focus from complex implementation (low-level source-coding) specific details to a design perspective (high-level system view). Modeling can incrementally translate functional aspects into service models, enabling service implementation through model-driven engineering or low-code development. The literature highlights that this modeling phase should utilise both existing and innovative notations, such as UML diagrams and quantum circuit models, to design quantum software services.}}

\vspace{0.5 em}

\scriptsize{~\textsf{\textbf{-- Patterns} complement software models by providing reuse knowledge and best practices of classical and quantum SE during service modelling. The review has identified several classical as well as quantum-specific patterns such as Quantum API Gateway and Qunatum-Classic Split that can facilitate architects to design quantum software services.}}
\end{tcolorbox}







\subsection{Service Assembly}
It refers to assembling the modeled services, i.e., to implement the designed services for their deployment and execution. Service assembly phase has two main activities namely \textit{service usecase} and \textit{implementation framework}. Service usecase identifies the practical scenario addressed or implemented by the service, whereas the implementation framework involves tools, technologies, and programming languages etc. that enable source-coding and execution of the service. As an example, the study [S18][S19] identifies quantum simulation as a usecase that is implemented using Flask, a python-based framework for microservice-based implementation of the simulation. This section discusses the identified usecases and implementation frameworks as part of this phase.

\subsubsection{Service Usecase}

The literature review has identified a total of 6 use cases that include varying scenarios ranging from quantum simulation to quantum data search and security. Service usecase represents a scenario that bridges the gap between service design (modeling) and its implementation (assembly), as a practical instance of conceived services. We introduce each of the usecase below:

\begin{itemize}
    \item \textit{Quantum simulations} is identified in a total of 4/41 studies, i.e., approximately 10\% of reviewed literature [S13][S19][S20][S35]. For example, the study [S13] proposes a framework for simulation of distributed quantum computing, i.e., quantum processors that are distributed across various nodes. The study enables an experiential development, deployment, and execution of quantum software services that simulate tasks on quantum processors to evaluate the functionality and performance of distributed quantum computing.

      \item \textit{Quantum search} is identified in a total of 2/41 studies, i.e., approximately 5\% of reviewed literature [S5][S17]. For example, the study [S5] develops an experimental solution to execute quantum microservives with a usecase on searching for the best path in the context of travelling salesman problem. Quantum microservice implementation is integrated into a hybrid microservices architecture, tested on Amazon Braket across three quantum hardware vendors. 
      
        \item \textit{Algorithmic execution} is identified in a total of 2/41 studies, i.e., approximately 5\% of reviewed literature [S6][S24]. Execution of classical or quantum specific algorithms, such as Shor's algorithm on quantum machines is a complex problem. For example , the study [S6] proposes a solution named quantum algorithm as a service enabling developers to abstract quantum components using software as a service and function as a service, supporting algorithm deployment across multiple QC providers and hardware platforms.

          \item \textit{Cyber Security} is identified in a total of 2/41 studies, i.e., approximately 5\% of reviewed literature [S23][S40]. For example, the study [S23] aims to develop services to provide secure access to QC resources, ensuring that users can reliably authenticate, authorize, and interact with quantum resources while maintaining data privacy and system integrity.
          
            \item \textit{Smart Systems} is identified in a total of 1/41 studies, i.e., approximately 2.5\% of reviewed literature [S11]. The study [S11] develops an experimental system to investigate quantum service-orientation in the context of smart city systems.

               \item \textit{Process Automation and Optimisation} is identified in a total of 4/41 studies, i.e., approximately 10\% of reviewed literature [S2][S5][S7][S15]. The study [S7] proposes a hybrid quantum–classical service approach using Amazon Braket and demonstrates its applicability through a software optimisation case study that tackles problems beyond the capabilities of classical algorithms.
               
            \item \textit{No explicit use case} is identified in a total of 16/41 studies, i.e., approximately 39\% of reviewed literature. For example [S14][S15] develop experimental systems but do not mention a specific usecase for implementation.
\end{itemize}

The type of use case can influence the selection of implementation frameworks, programming languages and tools for service implementation. For instance, study [S3] employs Q\#, within the Microsoft .NET framework to implement and execute quantum algorithms.

\subsubsection{Service Implementation Frameworks}
Quantum service implementation relies on frameworks and programming languages etc., which provide the notation and source code scripts necessary to develop, manage, and operationalise QC resources \cite{R15_de2022software} \cite{R21_perez2020towards}. In recent years, several Quantum Programming Languages (QPLs) such as Q\# (Microsoft) and Cirq (Google) have emerged, offering specialised syntax and execution environments for developing, executing, and deploying quantum source code. Analysing the programming languages helps determine whether classical languages (e.g., Python, Java, C) suffice for QCaaS implementation or if specialised QPLs are needed. The SMS results identify Python, Java, and Q\# as the primary languages for implementing quantum services. Figure \ref{Fig:Results} f) differentiates between native code and specialised libraries or APIs built using a specific language. For example, studies [S4][S5] used native Python code for quantum microservices experimentation, while study [S1] employed Flask, a Python-based web framework, to optimise quantum service delivery on Amazon Braket. Python remains the most widely used language, both in its native form and through frameworks and libraries such as Flask, Qiskit, and Google’s Cirq. For further details on quantum programming languages, a recently conducted study \cite{QPL} combines software repository mining and a practitioner survey as a mixed-method research approach to develop taxonomies of current uses and informs about emerging challenges in quantum programming as perceived by quantum source code developers.

\begin{itemize}
    \item \textit{Python} as a programming language represents the predominant choice to implement quantum software services identified in a total of 19/41 studies, i.e., 46\% of the total reviewed literature. Python-based service implementation can be generally classified into five sub-categories, such as native Python code [S4][S6][S35], Python Flask [S1][S5][S15], Python with Qiskit [S23][S29][S31], and Python with other languages and/or frameworks [S16][S20][S24]. For example, a quantum algorithm implemented in Python, can utilise the API Gateway pattern, for its deployment as quantum algorithm as a service on Amazon Braket platform [S6]. Python Flask is a lightweight web framework that can be used to develop web applications using Python. Flask as a web development framework for is considered as simple, flexible, and easy to get started with  making it a preferred option both beginners and professionals. For example, the source code for qunatum web services is generated in Python, using the Flask Web framework [S15].
    
    \item \textit{Qiskit} is an open-source QC framework that is developed by IBM. Qiskit is designed to support the creation, simulation, and execution of quantum algorithms \cite{QPL} and identified in a total of 4 studies [S11][S21][S25][S36]. Qiskit provides tools to design quantum circuits that are used for algorithm development to manipulate quantum hardware, enabling research and development in QC. For example, the study [S36]  presents a comparative analysis of qunatum software that is developed using Qiskit framework and deployed in a hybrid cloud environments.

    \item \textit{Microsoft’s platform languages} include several programming languages designed for developing quantum software and services that can be executed on Azure Quantum. The platform provides a comprehensive environment for quantum software development, supporting languages such as Q\# and C\#. Q\# is specifically tailored for quantum algorithm design, while C\# is used to handle classical logic and workflow orchestration. Together, they enable hybrid quantum-classical applications and seamless cloud-based execution, identified in a total of 2 studies [S3][S10]. For example, the study [S3] implements quantum function as a service for quantum application development in cloud environments using Q\#.

    \item \textit{Others} reprents service implementation using a language and/or implementation framework not covered in any of the above-mentioned languages, identified in one study [S12]. The study [S12] proposes OpenAPI is a specification for defining and documenting RESTful APIs in a standard, machine-readable format. It allows developers to describe an API’s endpoints, request/response formats, authentication, and other details, typically using JSON or YAML.
    
    \item \textit{No specific framerwork} refers to the studies that do not have any explicit mention of any programming languages, as shown in a total of 15 studies. For example, the study [S2] presents a conceptual model as a model-driven development of qunatum service computing, however, no details for any implementation have been provided.
    
\end{itemize}


\begin{tcolorbox} [sharp corners, boxrule=0.1mm,]
\faEdit \scriptsize{~\textsf{\textbf{Quantum service assembly} phase focuses on assembling, i.e., implementing the modelled services, that is mainly supported by:}}

\vspace{0.5 em}

\scriptsize{~\textsf{\textbf{Service usecase} representing the practical context or a usage scenario of quantum services that range from quantum search, to simulation, cyber security and proof-of-the-concept implementation of quantum algorithms. Usecass also impact the selection of programming languages and tools for implementation.}}

\vspace{0.5 em}

\scriptsize{~\textsf{\textbf{Service implementation} involves programming languages and/or frameworks as a system of notation for source-coding of quantum services. Classical programming languages, such as Python, represent a predominant
choice over recently emerged quantum programming languages (Q\#) due to more comprehensive documentation and familiarity with Python in the community of service developers.}}
\end{tcolorbox}

\subsection{Service Deployment}

As the final phase of the quantum service development lifecycle, the deployment phase focuses on selecting QC platforms for operationalising and executing assembled services \cite{ServiceDeploy2024}. Platform providers, or quantum vendors, supply both hardware and software infrastructure that enables service developers to build and utilise quantum services. Deployment can be represented using modeling notations such as UML deployment diagrams, as illustrated in Figure \ref{Fig:Pattern}. The SMS identified three primary quantum vendors for service deployment: \textit{Amazon Braket}, \textit{IBM Quantum}, and \textit{Rigetti}. Among these, Amazon Brake, a managed service on AWS, is the most preferred platform for designing, testing, and running quantum algorithms. Its popularity stems from its flexibility, allowing developers, including novices, to design their own quantum algorithms or leverage pre-built algorithms, tools, and documentation to develop and manage quantum services on the platform.

\begin{itemize}
    \item \textit{Amazon Braket} is a fully managed quantum computing service by AWS that enables users to design, simulate, and run quantum algorithms using multiple quantum hardware platforms and simulators, supporting research and hybrid quantum-classical workflows in a scalable cloud environment. Amazon Braket as a deployment platform is identified in a total of 12 studies such as [S1][S4][S5][S7]. For example, the study [S1] implements a Quantum API Gateway using Python and Flask on Amazon Braket, chosen for its unified support for executing quantum code across multiple quantum processors.

    \item \textit{IBM Quantum} is a cloud-based platform that provides access to quantum computers and simulators. It supports Qiskit for quantum programming, enabling users to design, test, and run quantum algorithms on real quantum hardware for research, education, and development. It has been identified in a total of 10 studies, such as [S3][S10][S14][S14][S18][S19][S20]. For example, the study [S19] uses a web-based frontend for users to build and run Qiskit-based applications, with a scheduler handling job queuing, backend selection, and result retrieval across multiple execution platforms.
    
    \item \textit{D-Wave Leap} developed by D-Wave systems is a quantum annealing computer designed to solve complex optimization problems. It uses quantum annealing rather than gate-based computation, making it suitable for tasks like logistics, machine learning, and scheduling on specialized quantum hardware. It has been identified in a total of 3 studies [S11][S23][S41]. The study [S41] presents a quantum-classic based cloud architecture as a blueprint that spans from code-level development to production-ready quantum software, implemented on the D-Wave quantum platform for solving real-world optimisation problems.
    
    \item \textit{Azure Quantum} is Microsoft’s cloud-based platform for quantum computing, offering access to diverse quantum hardware, simulators, and tools. It supports languages like Q\# and integrates with classical systems, enabling hybrid quantum-classical application development for research, education, and enterprise solutions. It has been identified in a total of 2 studies [S10][S34]. For example, the study [S10] utilises the Miscorost Azure platform to develop a prototype for qunatum search.
    
    \item \textit{Google Quantum AI} is Google’s initiative to advance quantum computing through hardware development, algorithm research, and open-source tools like Cirq. It focuses on achieving quantum advantage and enabling scalable, fault-tolerant quantum processors for scientific and industrial applications. It has been identified in a total of 2 studies [15][16]. For example, the study [S15] provides a technical comparison of quantum computing service providers using the Travelling Salesman Problem as a case study, implemented with Cirq. Empirical tests highlight key differences and suggest approaches to reduce vendor-specific limitations, such as vendor lock-in while utilising quantum software services.
    
    \item \textit{No Platform} is represented in a total of 11 studies. For example, the study [S12] provides a conceptual model that does not include any implementation and deployment of services. Such cases of conceptual solutions lack service implementation and do not provide any details on deployment platforms.
    
\end{itemize}

\begin{tcolorbox} [sharp corners, boxrule=0.1mm,]
\faEdit \scriptsize{~\textsf{\textbf{Quantum service deployment} is enabled by deploying the implemented services on a QC server - enabling quantum service-orientation. Deployment involves selecting quantum QC platforms for executing services, marking the final phase in the service lifecycle. This study identified Amazon Braket, IBM Quantum, and Rigetti as major platforms for quantum service deployment and delivery. Amazon Braket is the most preferred platform due to its accessible tools, support for custom algorithms, and ease of use for novice developers.}}
\end{tcolorbox}


\vspace{0.5em}

\section{A Reference Architecture for QCaaS}
\label{sec:RefArch}

In software engineering context, a reference architecture acts as a system-wide blueprint, defining computational elements as architectural components and representing their interactions through architectural connectors \cite{RefArchJSS-2011} \cite{RefArchSW-2015}. The component and connector level architectural view can enable software designers to apply reuse knowledge and best practices and facilitates communication between domain professionals (a.k.a. system stakeholders) to implement the architecture \cite{ABD1}. Based on the findings of RQ-1 presented in Section \ref{sec:RQ2SESolutions} and summarised in Table \ref{tab:DataExtraction}, we applied the layered architecture pattern \cite{ArchLayers-2005} to organise quantum service lifecycle activities into three layers: \textit{service development}, \textit{service deployment}, and \textit{service split}, as illustrated in Figure \ref{fig5:RefArch}. Figure \ref{fig5:RefArch} a) provides an abstract view to conceptualize the structural composition of the architecture, while Figure \ref{fig5:RefArch} b) translates this abstract view into a more detailed, concrete representation of the architecture in terms of i) architectural layers, ii) phases and activities encapsulated inside each layer, iii) human roles with tool support, and iv) service artifacts.  A fine-grained representation of the reference architecture, structured in individual layers and their encapsulated elements, is detailed below and visually represented in Figure \ref{fig5:RefArch}.   

\textit{Tool support for the reference architecture:} To support architecture-based development, we have developed a prototype named `QADL: Quantum Architecture Description Language' that provides a specification language, design space, and execution environment to architect quantum software services. QADL represents a proof-of-the-concept system that enables designers to specify architectural components (representing computational services), define component interconnections (enabling service communication) to support architecture-based development. Further details about QADL are provided in \cite{QADL25} 

\subsection{Layers of the Architecture}\label{sec:layers}

Architectural layering enables the separation of functional concerns, allowing elements that support similar functionality to be grouped within the same layer. For example, in Figure  \ref{fig5:RefArch}, the service development layer unifies the conception, modeling, and assembly of quantum services; the deployment layer covers service execution and hosting, and the service split layer handles the division between quantum and classical execution. Beyond structuring the system, a reference architecture provides a template for designing software solutions within a specific domain, in this case, quantum computing in general with a specific focus on quantum service-orientation \cite{RefArchJSS-2011}. It also establishes a common architectural vocabulary, linking functional requirements to architectural components and connectors. For example, within the service development layer, the Quantum Service Conception phase includes two key activities: \textit{Functional Specifications} and \textit{Quality Attributes}. These activities involve human roles, such as the service developer, who defines the required functionality and desired quality of the service. The outcome of this phase is a Quantum Significant Requirements (QSRs) artifact, which forms the foundation for service design. In summary, layered architectural structuring organises phases containing multiple activities, enabling human roles to incrementally produce service artifacts through conception, modeling, assembly, and deployment of QCaaS.

\subsection{Layering of Service Lifecycle Activities}\label{sec:phaseactivity}

The three architectural layers encompass the phases and activities of the service lifecycle, organized according to IBM’s service-oriented architecture (SOA) \cite{R20_keen2006patterns} to guide the development of the reference architecture. For example, in Table \ref{tab:DataExtraction}, the \textbf{Modeling} phase focuses on how to represent quantum services for implementation. Within this phase, the activities include: (i) \textit{modeling notation}, which uses UML-based class and component diagrams to enable service modeling, and (ii) \textit{patterns}, which provide best practices for modeling QC services. To better structure QCaaS, we split the `Model' activity from the SOA lifecycle into two distinct activities: `Conception' and `Modeling'. This separation distinguishes functional needs (conception) from their architectural representation (modeling). Conception defines the design specifications for the functional requirements of quantum services, while Modeling represents these requirements architecturally for implementation. For example, the first row in Table \ref{tab:DataExtraction} shows that the required functionality enabling quantum software delivery is architecturally represented using a UML deployment diagram and the API Gateway pattern. It is vital to mention that the reference architecture in Figure \ref{fig5:RefArch} omits the Manage phase from the SOA lifecycle, as the literature provides no evidence supporting identity, compliance, or business-metrics management for quantum services. The SMS process could not find any evidence in the literature that supports identity, compliance, and business metrics management of quantum services.

\subsection{Human Roles and Tools to Architect QCaaS}\label{sec:humanroles}
Human roles in the reference architecture capture the knowledge, expertise, and involvement required to develop or utilise quantum services \cite{R19_ahmadtowards}. Two primary types of roles are identified: \textit{service developers} and \textit{service users}. Service users may be individuals or teams within an organisation that require quantum computation. In contrast, service developers encompass various specialised roles, including quantum service developers, quantum algorithm designers, and quantum domain engineers. Recent studies emphasise the importance of quantum-specific expertise, such as quantum software architects, who can map QuBit operations to architectural components, and quantum code managers, who simulate and analyze quantum information flow \cite{X2_QSA}. A quantum domain engineer, for example, focuses on analysing quantum-specific attributes, such as the mapping between QuGates and their corresponding QuBit representations. Quantum domain engineers can provide input to the design and fabrication of quantum processing units (QPUs), optimising algorithms and programming languages to enhance performance. Quantum domain engineers also guide hardware and software teams in realizing Quantum Significant Requirements (QSRs), implementing them as quantum algorithms for execution on quantum computing platforms, as illustrated in Figure \ref{fig5:RefArch} b).

In addition to human roles, tool support plays a crucial role in enabling automation and user-driven customisation of architecting QCaaS \cite{QTool}. Tools such as Qiskit, Cirq facilitate circuit design, compilation, and optimisation, while platforms like IBM Quantum Composer, Amazon Braket, and Azure Quantum enable service orchestration and hybrid quantum–classical integration. These tools can enable automation of tasks such as QuBit utilisation, circuit design, resource allocation, and workflow execution to reduce manual intervention and improve efficiency of the architecting process \cite{QADL}. Tools can also provide customisation and adaptability through inputs from human roles such as quantum service developers, who define algorithms, and system architects, who specify service configurations and quality requirements. This synergy between tool-driven automation and human-provided input ensures that QCaaS development can be efficient with human oversight and supervision.

\subsection{Service Artifacts}\label{sec:artifacts}
Service artifacts are tangible outcomes such as documents, design models, or source code, that support the development and delivery of quantum services. In Figure \ref{fig5:RefArch}, each phase within the architectural layers produces a specific artifact. Four key artifacts are identified: QSRs, Quantum Service Design, Quantum Service Implementation, and Quantum Service Deployment. For instance, the Quantum Service Conception phase generates QSRs, capturing functional and quality attributes that guide the modeling and pattern-based design of the quantum service.

\begin{figure*}
\centering
\includegraphics[width=1.06\textwidth]{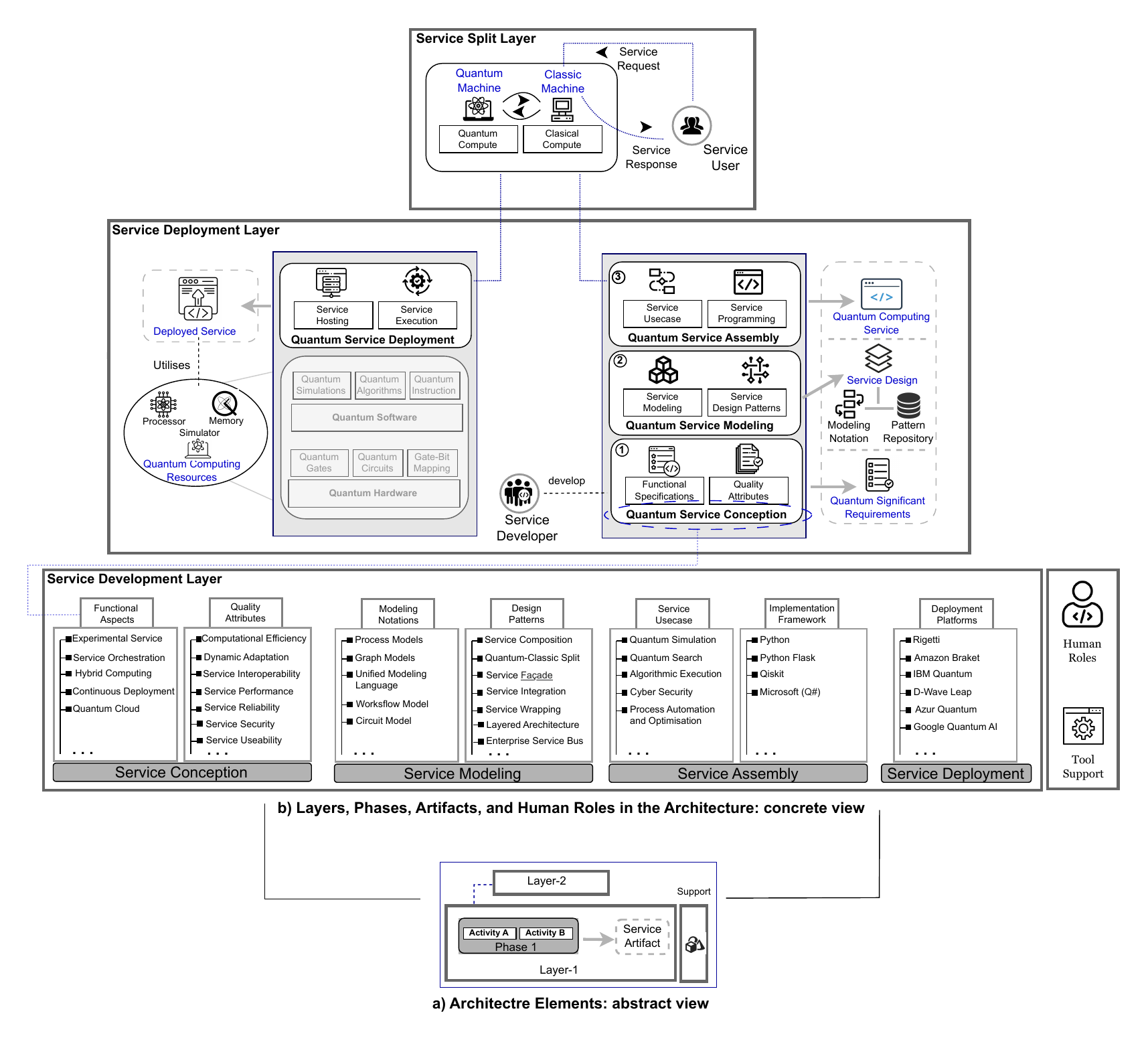}
\caption{\label{fig5:RefArch} Reference Architecture for QCaaS - a Layered View.}
\end{figure*}

\subsection{Example Scenario for the Reference Architecture}
\label{sec:implement}

The example scenario represents a QCaaS usecase that is used to exemplify each phase and its corresponding activity in the reference architecture. The scenario does not reflect a comprehensive or full-scale development of the solution, it only elaborates a usecase of the reference architecture that provides the basis for further implementation and validation of the architecture. The example scenario as a specific instance of implementation is explained later in this section while discussing the service assembly. Implementation of the usecases is supported by QADL and detailed in our previous work \cite{QADL25}.



\begin{itemize}
    \item \textit{Phase I - Service Conception:}\label{sec:conception}
The listings below outline the functional specifications for quantum services designed to generate the prime factors of a given integer. To achieve this functionality, computations are divided between a classical compute and a quantum compute. Additionally, it is necessary to evaluate both the correctness of the implementation and the efficiency of the prime factorisation solution. The listings capture the functional aspects alongside two quality attributes, focusing on the efficient utilisation of QC resources, particularly the effective use of QuBits. Quality attributes complement the functional specifications by defining the desired characteristics and performance standards of the solution.

\begin{tcolorbox}[colback=white]
Design solution for inputting the integer \textsf{N} and outputting prime factors \textsf{F}. 

\vspace{0.2em}
\textsf{- Split} the solution for classical and quantum hybrid computations.

\vspace{0.2em}
\textsf{- Validate} correctness of solution and efficiency of computations.
\end{tcolorbox}

\begin{tcolorbox}[colback=white]
\textsf{QuBit utilisation} The designed solution to efficiently utilise QuBits used for factoring integers of the given size.
\end{tcolorbox}

For example, Figure \ref{fig6:Implementation} illustrates two architecting activities within the service conception phase, focusing on specifying the functional and quality aspects of prime factorisation. These aspects are captured as Quantum Significant Requirements (QSRs), serving as an architectural artifact of the conception phase. Service architects and developers can use the QSRs to apply appropriate modeling notations (e.g., UML, ADL) and patterns (e.g., service orchestrator) to create a service design, which acts as an artifact for assembling the microservices.





\vspace{0.5em}

\item \textit{Phase II - Service Modeling:} \label{sec:modeling}
The QSRs, produced during the conception phase, serve as the foundation for creating a service design model that reflects the functional aspects of the quantum service. The quantum service modeling phase builds on this foundation through the following two activities.


\vspace{0.3 em}
\textit{A. Modelling notation for quantum services:} 

As shown in Figure \ref{fig6:Implementation}, UML is used as the modeling notation to create the quantum service model, employing UML component and sequence diagrams. To support quantum-enabled modeling in QSE, the quantum UML profile extends traditional UML diagrams for both structural and behavioral modeling of quantum software \cite{QUML}. The UML component diagram in Figure \ref{fig6:Implementation} provides a structural view of quantum services and their interconnections. According to the UML profile \cite{UMLSOSYM2025}, each service is represented as a separate component that offers a computational function (e.g., generating a random number) and communicates with other components via connectors. For instance, the \textsf{GetGCD} service connects with the Controller service to compute the Greatest Common Divisor (GCD) of a randomly generated number as part of the algorithm. In contrast, the UML sequence diagram captures the behavioral view, illustrating message passing between services to enable communication. For example, in the sequence diagram, the controller service \textsf{C:Controller} sends a Generate(N) message to the number generator service \textsf{N:NumGenerator}, which generates a random number R and returns it to the controller. Other UML diagrams, such as use case diagrams, can also represent QSRs. For demonstration, we used component and sequence diagrams to exemplify service modeling, highlighting both the structural composition and behavioral interactions of services to be assembled.

\vspace{0.3 em}
\textit{B. Pattern-based modelling:}  Service design and development patterns provide reusable knowledge and best practices for engineering service-oriented solutions \cite{R20_keen2006patterns}. In Figure \ref{fig6:Implementation}, we applied two patterns: \textit{Orchestrator} and \textit{Quantum-Classic Split}. The Orchestrator pattern, a standard SOA pattern, manages the execution of multiple services to complete a service-driven task. For example, it coordinates the factorisation of a randomly generated prime number through the \textsf{NumGenerator} and \textsf{Factorise} services. The Classic-Quantum Split pattern divides functionality between classical and quantum computing, enabling hybrid quantum computing. In Figure \ref{fig6:Implementation}, the Controller service orchestrates operations between classical and quantum microservices: \textsf{NumGenerator} and \textsf{GetGCD} run on a classical machine, while \textsf{QuantumModularExponentiation}, \textsf{QuantumInverseQFT}, and \textsf{Factorise} execute on the quantum machine. Within the reference architecture, the modeling phase leverages both modeling notations and patterns to create a service design -- a visual model that can be assembled into executable quantum microservices.

\vspace{0.5 em}

\item \textit{Phase III - Service Assembly:}\label{sec:assembly}
Service assembly involves assembling a service by identifying a use case and developing the corresponding program that implements it, shown in Figure \ref{fig6:Implementation} and described below.

\vspace{0.3em}
\textit{A. Service usecase:}
The implementation details based on Figure \ref{fig6:Implementation}, which illustrates the conception, modeling, assembly, and deployment of software services implementing Shor’s algorithm. Shor’s algorithm is a quantum computing algorithm for factoring integers in polynomial time \cite{Shor2024}, with applications including prime factorisation, cryptography (e.g., breaking RSA), and period finding for functions. In this study, we focus specifically on developing microservices that run on a quantum computing platform to implement Shor’s algorithm for prime factorisation. Figure \ref{fig6:Implementation} provides an illustrative example of how services—modeled using the reference architecture in Figure \ref{fig5:RefArch} can be developed to realize Shor’s algorithm. The reference architecture in Figure \ref{fig5:RefArch} serves as a blueprint, outlining the necessary steps and structure to architect a quantum service-oriented solution.

\textit{B. Service implementation framework:}
To implement the use case, the service must be programmed for execution, meaning the design specifications from the modeling phase are translated into executable specifications during service assembly. Figure \ref{fig6:Implementation} illustrates this process, showing both the algorithmic design and its implementation as part of service assembly. The example demonstrates a partial assembly of services, including classical services that generate a random number and compute the GCD, which support the quantum computation in Shor’s algorithm. A UML sequence diagram serves as the model to guide assembly, mapping the algorithm to its implementation in a programming language. A C\# code snippet provides a summarised view of the \textsf{PrimeFactorisationController} service, which orchestrates both classical and quantum services. Figure \ref{fig6:Implementation} effectively demonstrates how the reference architecture can be applied in practice, providing a concrete example of implementing quantum service-orientation.

\item \textit{Phase IV - Service Deployment} As the final phase in the reference architecture, deployment comprises of service hosting and service execution. Both activities largely depend on the quantum computing platform where the quantum services being deployed.


\textit{Service Hosting Platform:}
The hosting of the assembled services is depicted using a UML deployment diagram in Figure \ref{fig6:Implementation}. The diagram shows two deployment nodes representing the hosting machines: the classical compute node hosts the \textsf{NumGenerator}, \textsf{GetGCD}, and \textsf{Controller} services, while the quantum compute node hosts \textsf{QuantumModularExponentiation}, \textsf{QuantumInverseQFT}, and \textsf{Factor} services. The hosting of the assembled service is presented as a UML deployment diagram as in Figure \ref{fig6:Implementation}. UML deployment diagram show two deployment nodes (hosting machines). The classical compute node hosts three services \textsf{NumGenerator}, \textsf{GetGCD}, and \textsf{Controller}, whereas the quantum compute node hosts three services, namely \textsf{QunatumModularExponentiation}, \textsf{QunatumInverseQFT}, and \textsf{Factor}.

\textit{Tool support and architectural language:} Figure \ref{fig6:Implementation} b) shows the user interface of QADL, a prototype that we developed to model and assemble, i.e., represent and formally describe quantum software components that can provide computational services. Drawing inspiration from classical architectural description languages (ADLs) \cite{ADLClass}, the quantum architecture description language (QADL) incorporates (a) a graphical environment for modeling software components, (ii) automated parsing for syntactic verification of the components, and (iii) execution of compoent services through integration with IBM Qiskit. In Figure \ref{fig6:Implementation} b), left-side, the QADL script specifies the architectural description, declaring QuBits and applying the QUGates (Hadamard, Pauli-X, and CNOT), followed by measurement. This description corresponds to the core quantum subroutine of Shor’s algorithm - superposition (via Hadamard), controlled operations for modular exponentiation (via CNOT), and state inversion (via Pauli-X) - leading to the measurement for computing the periodicity.

\begin{landscape}
\begin{figure}
\centering
\includegraphics[width=1.43\textwidth]{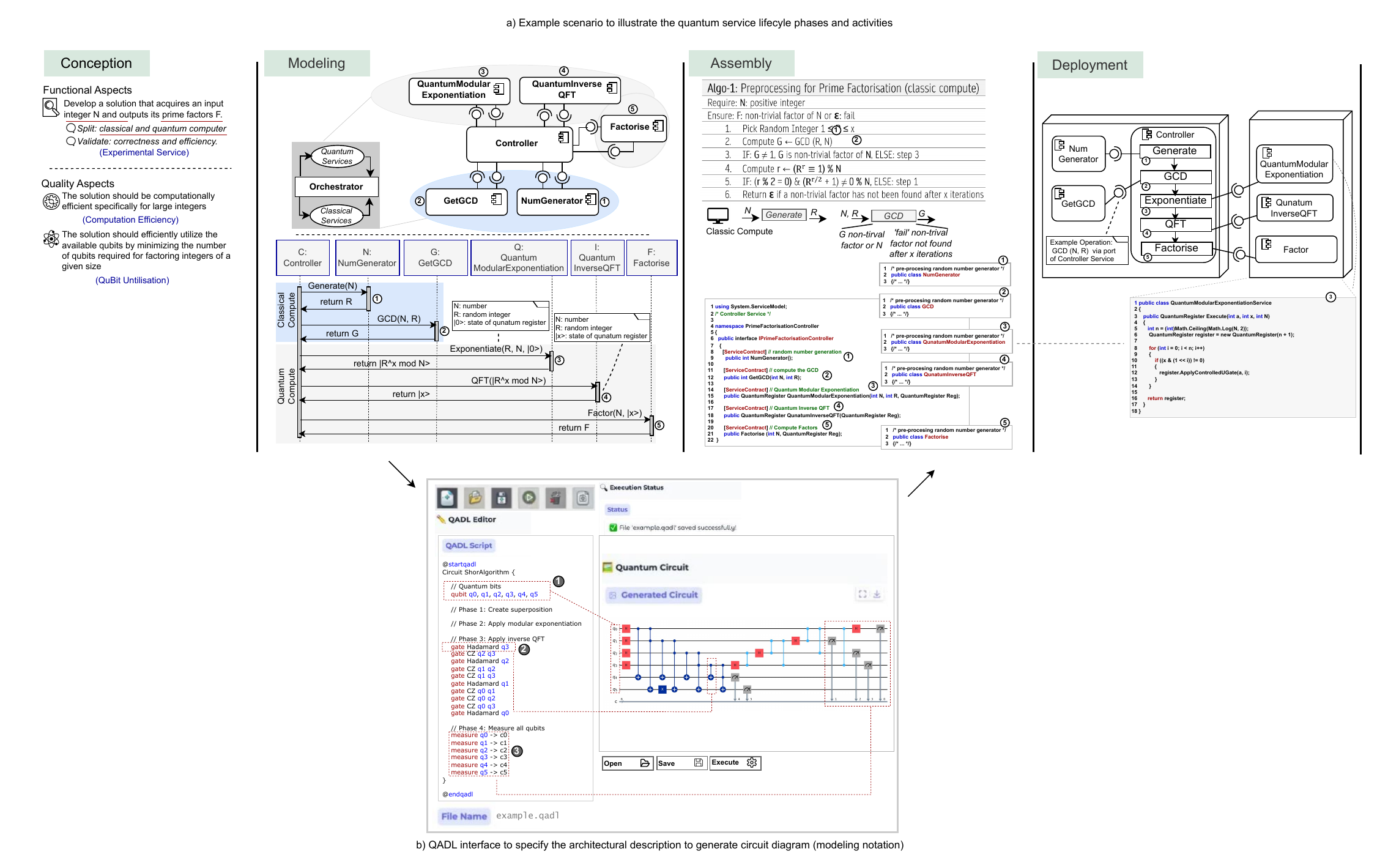}
\caption{\label{fig6:Implementation} Reference Architecture Based Representation of the Quantum Service Lifecycle.}
\end{figure}
\end{landscape}

QADL script for architectural specification of Shor's algorithm via a quantum circuit diagram, as in Appendix \ref{sec:appendix} - Listing \ref{lst:QADLScript}.

In Figure \ref{fig6:Implementation} b), towards the right, QADL represents qunatum circuit diagram to model the algorithm. The description and representation with QADL enables Shor’s algorithm to be modeled, visualised, and executed as a service abstraction that supports design at higher-level of software abstraction rather than low-level implementations. Low-level details, such as circuit creation and component execution within QADL is enabled via IBM Qiskit, an open-source quantum software development framework created by IBM \cite{IBMQISKIT}. The Qiskit provides the tools to design, simulate, and run quantum programs on both simulators and real quantum hardware, its primary role is to provide a modular and descriptive framework for describing quantum software architectures.

\end{itemize}

\section{Emerging Trends of QCaaS (RQ-3)}
\label{sec:RQ3Emerging}
We now answer RQ-3 that aims to report the emerging trends, as dimensions of potential future research, on engineering software services for quantum computing. To identify these trends, we specifically reviewed the details about the objectives, evaluation/demonstrations, as well as limitations and future research details in each reviewed study.  This helped us to identify a number of trends that can be generally classified into five (5) main areas, detailed in this section.  The identified trends presented in this section do not represent a comprehensive list, instead, it highlights some possible future research based on the data from existing literature. To maintain a consistent presentation of results, we have visualised the emerging trends identified in Figure \ref{FutureResearch} according to the service lifecycle activities \cite{R7_bouguettaya2017service} \cite{R20_keen2006patterns} as in Table \ref{tab:DataExtraction}. For example, Figure \ref{FutureResearch} highlights that during the conception of quantum services, an empirical identification and thematic classification of QSRs is an area of future research. QSRs include quality aspects that complement the functional aspects to ensure the required functionality and desired quality of QCaaS as part of quantum domain engineering activity in QSE. 



\subsection{Process and Human Support for QCaaS} 

Process-centred engineering is reflected in the service lifecycle, which structures multiple activities, such as service design, development, and delivery, into a coherent, iterative, and incremental process \cite{R19_ahmadtowards}. Although the reviewed studies organise their findings under the SOA life cycle, most lack an explicitly process-centric approach, thereby underscoring the need for processes tailored to QSE \cite{R19_ahmadtowards}. Emerging paradigms such as quantum DevOps, quantum microservicing, and agile methods for quantum services/software represent quantum-specific adaptations of established software engineering processes that align with the unique requirements of QSE \cite{R23_khan2022agile}. For example, in an agile approach to QSE, quantum domain engineering can capture domain-specific knowledge of quantum systems (both hardware and software) to develop design models that serve as blueprints for implementing quantum software and services. Process-centric approaches can also support tools (for automation) and professional roles (human decision support) to engineer quantum services. We identified the need for human roles as QSE professionals to effectively undertake QCaaS development. Consequently, there is a clear need to define and develop QSE specific professionals expertise such as quantum architects, domain engineers, and service developers, capable of effectively executing Quantum Computing as a Service (QCaaS) development processes.

\vspace{0.3em}

\faComment
~ \small{\textit{Process-centred quantum service-orientation} can enable service developers to encapsulate qunatum source code and algorithms as modular and reusable services, designed for composition, orchestration, and execution within process-driven workflows. It can enable the assembly of end-to-end services, e.g., for optimisation, simulation, or secure cryptography, thereby shifting the focus from isolated quantum algorithms to holistic, process-aware quantum solutions.}

\vspace{0.3 em}

\faComment
~ \small{\textit{Human roles}} play a pivotal part in enriching the process-centric approach by synergising the theoretical knowledge of quantum physics with the practical disciplines of software engineering. This interdisciplinary integration enables effective support for key activities such as quantum domain engineering, software architecting, and simulation management, areas that currently lack sufficient expertise within quantum service-orientation. By integrating human expertise across these process-centered activities, the architecting process can better address the complexity of designing, developing, and managing quantum-enabled services, ultimately bridging the gap between quantum theory and software practice.}




\subsection{Empiricism for Managing the QSRs}  

The concept of Quantum Significant Requirements (QSRs) builds upon the well-established notion of Architecturally Significant Requirements (ASRs) in traditional software and service engineering \cite{ASR}. Within quantum domain engineering, QSRs play a crucial role in eliciting and documenting both functional and non-functional (quality) aspects of quantum software systems. While existing research has largely concentrated on the functional dimensions of quantum services, it has often overlooked critical non-functional aspects such as QuBit utilisation, energy efficiency, quantum vendor lock-in, QPU elasticity, and quantum error mitigation. These quality attributes, however, are fundamental as they explicitly influence the design, development, and operationalisation of QCaaS solutions. For example, during the quantum domain engineering activity, the hardware aspects (such as the operations of QuGates) can be systematically mapped to software aspects (such as components and connectors). This mapping enables the splitting of computational tasks between classical and quantum computers by applying the classic–quantum split pattern, thereby ensuring an efficient allocation of processing responsibilities across the hybrid computing environment.

Empirical analysis of available data available via the social coding platforms (e.g., GitHub) grounds theoretical QSRs identified via literary analysis, like Qubit utilisation and error mitigation, in a practical and developer-centric context. By examining real-world projects, such as those on GitHub, research can validate or refine the identified requirements with actual developer priorities. This evidence-based approach transforms QSRs from abstract concepts into a practical taxonomy reflective of real quantum software development needs \cite{QuantumMeetsSE}. Furthermore, empirical identification and classification of the QSRs can lead to the creation of a comprehensive QSR taxonomy. By combining the findings of the literature review with data mined from open-source repositories can capture both the existing and emergent, implicit requirements from developer artifacts like issue trackers of qunatum software systems. Classifying these findings allows for the development of a taxonomical classification of QSRs and identifies critical trade-offs, such as between quantum circuit depth and quantum error mitigation \cite{R19_ahmadtowards}.

\vspace{0.3 em}

\faComment
~ \small{\textit{Quantum significant requirements} are rooted in the concept of Architecturally Significant Requirements (ASRs), which aim to identify, classify, and document the functional and quality aspects of quantum software services. QSRs for quantum services extend this notion by complementing functional requirements with quality attributes, ensuring that both the required functionality and the desired quality of service are achieved. Furthermore, empirical mining of open-source and social coding platforms can help validate these theoretically derived QSRs, transforming them into a practical, evidence-based taxonomy that reflects the real-world development challenges encountered in quantum software engineering.}

\subsection{Model-driven Quantum Software Servicing}
Model-Driven Service Engineering (MDSE) empowers software engineers and architects to use models as high-level abstractions that hide complex, implementation-specific details while providing human-comprehensible visual notations for designing and implementing software services. Through model transformations, MDSE enables the conversion of design models into implementation artifacts (e.g., source code) and validation models (e.g., test cases). Modeling notations such as the Service-Oriented Architecture Modeling Language (SoaML) and the Quantum UML (Q-UML) specification \cite{QUML} extend traditional modeling approaches by offering metamodels and UML profiles tailored for service specification and design within service-oriented architectures. In the context of quantum software engineering, MDSE can particularly aid novice developers by helping them map algorithmic flows and source code modules to graphical representations, thereby supporting low-code, model-driven development of quantum services.

\vspace{0.3 em}

\faComment
~ \small{\textit{MDSE applies model-driven engineering principles to quantum computing by using notations like SoaML and Q-UML. It abstracts complex implementations into design-level models driven by Quantum Significant Requirements (QSRs). MDQSD enables automation through model transformations and supports human decision-making via tool-assisted design, promoting efficient and low-code development of quantum services.}

\begin{figure*}[]
 \centering
 \includegraphics[scale=0.8]{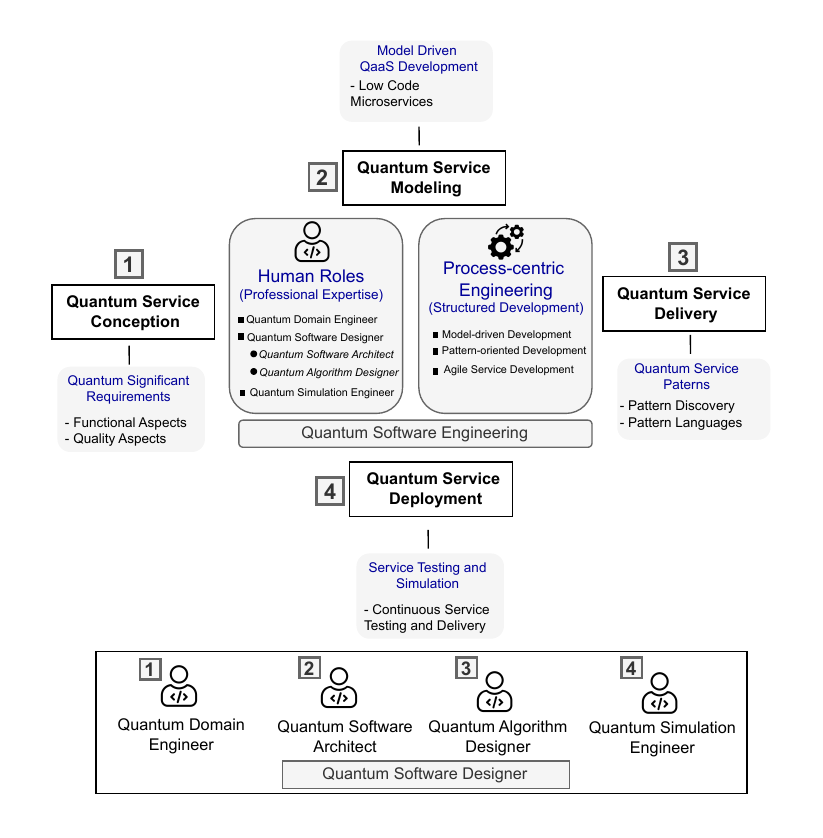} 
 	\caption{Overview of Emerging Research Trends}
	\label{FutureResearch}
\end{figure*}


\subsection{Discovery of Quantum Service Patterns}

Pattern-based software service engineering leverages architectural designs, implementation strategies, and best practices that can be reused to deliver software services \cite{R20_keen2006patterns}. While existing solutions employ patterns such as the quantum-classic split and service wrapping, there is little empirical evidence on discovering new patterns and tactics as reusable knowledge \cite{R27_GitQSE}. This lack of empiricism limits the reusability of design and implementation insights in quantum service engineering. One promising approach for pattern discovery is mining software repositories or social coding platforms (e.g., GitHub) to extract reusable design knowledge. Pattern-based solutions can complement human expertise, enabling novice developers to abstract the complexities of quantum source code while modularising solutions through well-established patterns.

\vspace{0.3 em}

\faComment
~ \small{\textit{Empirical Discovery of Patterns} Empirically grounded methods, such as mining repositories or social coding platforms for quantum software, can capture reusable design rationale for quantum service engineering. Pattern languages enable quantum software architects and algorithm designers to rely on proven best practices rather than ad-hoc, one-off solutions.}


\subsection{Continuous Testing and Delivery of Services}

With the adoption of agile methods in quantum software development \cite{R23_khan2022agile}, there is a need for light and adaptive approaches to enable continuous development and delivery of quantum services. The literature highlights a gap in testing quantum services. Continuous testing and delivery (CT/CD) can help validate services against QSRs more effectively and accelerate their deployment. Due to quantum-specific characteristics like superposition and entanglement, traditional testing methods are often insufficient, making specialised testing strategies essential to ensure the reliability and performance of quantum services. Furthermore, integrating CT/CD practices into the quantum software development lifecycle facilitates early detection of issues, reduces feedback times, improves contonuous delivery of QC services \cite{QDevops}.

\vspace{0.3 em}

\faComment
~ \small{\textit{Continuous Testing and Delivery} (CT/CD) grounded in agile methods, enables the rapid and reliable delivery of quantum software services. It also offers strategic benefits to vendors by facilitating the addition of new services to their quantum platforms.}

\vspace{0.3 em}

\textit{Large Language Models and Quantum Service Engineering}: Although we did not find any explicit evidence, specifically published research on synergising generative artificial intelligence (including the large language models (LLMs)) and quantum service development. However, some recent research studies indicate a growing trend on exploiting LLMs to design and develop quantum software. Recent work shows that LLMs can mimic the behaviour of quantum circuits and predict how quantum states evolve, thus enabling a faster and resource efficient alternative to traditional quantum simulators \cite{LLM-QSim25}. Purpose-built LLMs such as the GroverGPT, have claimed to achieve near-perfect accuracy when replicating Grover’s search algorithm on systems with up to 20 qubits \cite{GroverGPT-24}. Beyond the general purpose simulations, LLM-based solution are improving the quality of quantum source code through tools such as LintQ-LLM, which can detect, locate, and explain programming errors in Qiskit more effectively than standard static analysis methods \cite{LLM-QPL-25}. They are also being used to help developers keep up with rapid changes in quantum programming libraries, such as migrating between Qiskit versions, by automatically identifying and classifying common refactoring scenarios \cite{LLM-QRefactor-25}. The collective impact of these advancements indicates that LLMs, already recognised for their role in software engineering and architecting \cite{EASEGPT}, are likely to become central to quantum software engineering - enabling code generation and testing, as well as supporting the simulation, development, and maintenance of quantum software services.

\faComment
~ \small{\textit{LLMs for Quantum Service Engineering:} Recent studies show LLMs can simulate quantum circuits, enhance code quality, and support library migration in quantum software development. While direct research on combining LLMs with quantum service computing is scarce, emerging tools like GroverGPT and LintQ-LLM indicate LLMs may soon become central to QSE.}


\section{Related Surveys and Empirical Studies}\label{sec:Related}

We now review the most relevant existing research in terms of survey-based research on (i) quantum software engineering (Section \ref{Related:QSE}), and (ii) quantum services computing (Section \ref{Related:QAAS}). The discussion in this section is complemented with  data in Table \ref{tab:RelatedWork} that acts as a structured catalogue for criteria-based comparative analysis of proposed vs existing research. 

\textit{Interpretation of the Comparative Analysis:} For an objective comparison, Table \ref{tab:RelatedWork} presents each study using five-point self-explanatory criteria including (a) type of study (adopted from \cite{R16_petersen2008systematic}), (b) focus of study, (c) core findings, (d) QSE activity supported by the study, and (e) year of publication, each exemplified below. For example, the study \cite{X2_QSA} presents a systematic review, conducted in 2021, as part of evidence-based software engineering to focus on architecting quantum software. The SLR presents the core findings about architectural life cycle and state-of-the-art on architectural modeling, patterns, and tools to architect quantum software. The SLR focuses on design and architecting activities of QSE life cycle and indicates the needs for future research on quantum-specific professional expertise, i.e., human roles in QSE.

\begin{table*}[hbt!]
\caption{Summary of the Core Findings of Most Relevant Secondary Studies}
\begin{centering}
{\tiny
\begin{tabular}{|lcclcl|}
\hline
\rowcolor[HTML]{DAE8FC} 
\multicolumn{6}{|c|}{\cellcolor[HTML]{DAE8FC}\textbf{Quantum Software Engineering}} \\ \hline
\rowcolor[HTML]{EFEFEF} 
\multicolumn{1}{|l|}{\cellcolor[HTML]{EFEFEF}\textit{\begin{tabular}[c]{@{}l@{}}Study \\ Reference\end{tabular}}} & \multicolumn{1}{c|}{\cellcolor[HTML]{EFEFEF}\textit{\begin{tabular}[c]{@{}c@{}}Type of \\ Study  \end{tabular}}} & \multicolumn{1}{c|}{\cellcolor[HTML]{EFEFEF}\textit{\begin{tabular}[c]{@{}c@{}}Focus of \\ Study\end{tabular}}} & \multicolumn{1}{c|}{\cellcolor[HTML]{EFEFEF}\textit{\begin{tabular}[c]{@{}c@{}}Core\\ Findings\end{tabular}}} & \multicolumn{1}{l|}{\cellcolor[HTML]{EFEFEF}\begin{tabular}[c]{@{}l@{}}QSE\\ Activity\end{tabular}} & \begin{tabular}[c]{@{}l@{}}Publication \\ Year\end{tabular} \\ \hline

\multicolumn{1}{|l|}{\cite{X1_QSE}} & \multicolumn{1}{c|}{\begin{tabular}[c]{@{}c@{}}Literature \\ Review\end{tabular}} & \multicolumn{1}{c|}{\begin{tabular}[c]{@{}c@{}}Quantum \\ Software \\ Engineering\end{tabular}} & \multicolumn{1}{l|}{\begin{tabular}[c]{@{}l@{}}QSE life cycle, quantum software \\ engineering processes, methods, and tools.\end{tabular}} & \multicolumn{1}{c|}{\begin{tabular}[c]{@{}c@{}}Software  \\ Life cycle\end{tabular}} & 2020 \\ \hline

\multicolumn{1}{|l|}{\cite{X2_QSA}} & \multicolumn{1}{c|}{\begin{tabular}[c]{@{}c@{}}Systematic \\ Review\end{tabular}} & \multicolumn{1}{c|}{\begin{tabular}[c]{@{}c@{}}Quantum\\ Software \\ Architecture\end{tabular}} & \multicolumn{1}{l|}{\begin{tabular}[c]{@{}l@{}}Quantum software architecting activities, \\ modelling notations, patterns, and tools \\ for architectural development\end{tabular}} & \multicolumn{1}{c|}{\begin{tabular}[c]{@{}c@{}}Software Design \\ and \\ Architecture\end{tabular}} & 2021 \\ \hline

\multicolumn{1}{|l|}{\cite{R15_de2022software}} & \multicolumn{1}{c|}{\begin{tabular}[c]{@{}c@{}}Repository Mining\\ and   \\ Practitioner Survey\end{tabular}} & \multicolumn{1}{c|}{\begin{tabular}[c]{@{}c@{}}Quantum\\ Programming\\ Languages\end{tabular}} & \multicolumn{1}{l|}{\begin{tabular}[c]{@{}l@{}}Mining repositories and interviewing \\ practitioner to investigate quantum \\ programming languages\end{tabular}} & \multicolumn{1}{c|}{\begin{tabular}[c]{@{}c@{}}Software \\ Implementation\end{tabular}} & 2022 \\ \hline

\multicolumn{1}{|l|}{\cite{x4_Issues}} & \multicolumn{1}{c|}{\begin{tabular}[c]{@{}c@{}}Repository \\ Mining\end{tabular}} & \multicolumn{1}{c|}{\begin{tabular}[c]{@{}c@{}}Quantum\\ Issues\end{tabular}} & \multicolumn{1}{l|}
{\begin{tabular}[c]{@{}l@{}}Mining repositories to identify technical\\ debts in open-source quantum software\end{tabular}} &  \multicolumn{1}{c|}{\begin{tabular}[c]{@{}c@{}}Software \\ Implementation \\ and   Testing\end{tabular}} & 2022 \\ \hline

\multicolumn{1}{|l|}{\cite{X5_QC}} & \multicolumn{1}{c|}{\begin{tabular}[c]{@{}c@{}}Vision \\  Paper\end{tabular}} & \multicolumn{1}{c|}{\begin{tabular}[c]{@{}c@{}}Quantum\\ Software \\ Architecture\end{tabular}} & \multicolumn{1}{l|}
{\begin{tabular}[c]{@{}l@{}}QSE life cycle and comparing   quantum\\ computers with their classical \\ counterparts and vision for future research\end{tabular}} &  \multicolumn{1}{c|}{\begin{tabular}[c]{@{}c@{}}Software  \\ Life cycle\end{tabular}} & 2022 \\ \hline

\multicolumn{1}{|l|}{\cite{QuantumFrontier2024}} & \multicolumn{1}{c|}{\begin{tabular}[c]{@{}c@{}}Systematic Mapping\\ Study\end{tabular}} & \multicolumn{1}{c|}{\begin{tabular}[c]{@{}c@{}}Quantum\\ Software \\ Engineering\end{tabular}} & \multicolumn{1}{l|}
{\begin{tabular}[c]{@{}l@{}}Mapping QSE research, identifying \\key topics, study types, results,\\ popular quantum tools, and the \\evolution of community interest\end{tabular}} &  \multicolumn{1}{c|}{\begin{tabular}[c]{@{}c@{}}Facilitate knowledge transfer\\ (i.e., development and evolution,\\ the tools and frameworks \\being studied, etc.)\end{tabular}} & 2024 \\ \hline

\multicolumn{1}{|l|}{\cite{SECQSE}} & \multicolumn{1}{c|}{\begin{tabular}[c]{@{}c@{}}Literature \\Review\end{tabular}} & \multicolumn{1}{c|}{\begin{tabular}[c]{@{}c@{}}Security challenges\\ associated \\with QSE\end{tabular}} & \multicolumn{1}{l|}
{\begin{tabular}[c]{@{}l@{}}Addressing security in quantum\\ programming languages\\, hardware limitations, and\\ their impact on software \\development, highlighting vulnerabilities \\and advocating for early integration\\ of security measures to prevent exploitation.\end{tabular}} &  \multicolumn{1}{c|}{\begin{tabular}[c]{@{}c@{}}Software Design and\\ Testing\end{tabular}} & 2023 \\ \hline

\multicolumn{1}{|l|}{\cite{ChallengesQSE}} & \multicolumn{1}{c|}{\begin{tabular}[c]{@{}c@{}}Literature \\Review\end{tabular}} & \multicolumn{1}{c|}{\begin{tabular}[c]{@{}c@{}}Quantum Software\\ Engineering\end{tabular}} & \multicolumn{1}{l|}
{\begin{tabular}[c]{@{}l@{}}Investigating the evolving \\landscape of QSE, synthesising\\ insights to highlight key\\ challenges and the need\\ for robust frameworks and \\methodologies to advance the of QSE.\end{tabular}} &  \multicolumn{1}{c|}{\begin{tabular}[c]{@{}c@{}}Software\\ Lifecycle\end{tabular}} & 2024 \\ \hline


\rowcolor[HTML]{DAE8FC} 
\multicolumn{6}{|c|}{\cellcolor[HTML]{DAE8FC}\textbf{Quantum Services Computing}} \\ \hline

\multicolumn{1}{|l|}{\cite{R11_leymann2020quantum}} & \multicolumn{1}{c|}{\begin{tabular}[c]{@{}c@{}}Literature \\ Survey\end{tabular}} & \multicolumn{1}{c|}{\begin{tabular}[c]{@{}c@{}}Quantum \\ Cloud \\ Computing\end{tabular}} & \multicolumn{1}{l|}{\begin{tabular}[c]{@{}l@{}}Programming quantum computers and \\ investigating hybrid software consisting \\ of classical parts  and quantum parts.\end{tabular}} & \multicolumn{1}{c|}{\begin{tabular}[c]{@{}c@{}}Service Design \\ and \\ Architecture\end{tabular}} & 2020 \\ \hline

\multicolumn{1}{|l|}{\cite{QCCChallenge}} & \multicolumn{1}{c|}{\begin{tabular}[c]{@{}c@{}}Literature \\ Review\end{tabular}} & \multicolumn{1}{c|}{\begin{tabular}[c]{@{}c@{}}Quantum \\ Cloud \\ Computing\end{tabular}} & \multicolumn{1}{l|}{\begin{tabular}[c]{@{}l@{}}Reviewing advances, challenges, \\and future directions in \\quantum cloud computing, \\covering models, platforms,\\ technologies, and key\\ issues (e.g., resource\\ management, security, and privacy). \\Identifying open problems\\ and highlights opportunities\\ for future research.\end{tabular}} & \multicolumn{1}{c|}{\begin{tabular}[c]{@{}c@{}}Service Lifecycle\end{tabular}} & 2024 \\ \hline

\multicolumn{1}{|l|}{\cite{DQC}} & \multicolumn{1}{c|}{\begin{tabular}[c]{@{}c@{}}Literature \\ Review\end{tabular}} & \multicolumn{1}{c|}{\begin{tabular}[c]{@{}c@{}}Distributed Quantum\\ Computing\end{tabular}} & \multicolumn{1}{l|}{\begin{tabular}[c]{@{}l@{}}Discussing the challenges and open\\ issues in distributed quantum computing \\from a computing and communications\\ perspective.\end{tabular}} & \multicolumn{1}{c|}{\begin{tabular}[c]{@{}c@{}}Distributed Quantum\\ Computing (Algorithms, Compiling,\\ Networking
and\\ Simulation)
\end{tabular}} & 2024 \\ \hline

\multicolumn{1}{|l|}{\cite{x7_QAAS}} & \multicolumn{1}{c|}{\begin{tabular}[c]{@{}c@{}}Literature\\ Survey\end{tabular}} & \multicolumn{1}{c|}{\begin{tabular}[c]{@{}c@{}}Quantum \\ Service \\ Computing\end{tabular}} & \multicolumn{1}{l|}{\begin{tabular}[c]{@{}l@{}}Potential and limitation  of integrating \\ quantum computing with cloud computing\end{tabular}} & \multicolumn{1}{c|}{\begin{tabular}[c]{@{}c@{}}Service \\ Lifecycle\end{tabular}} & 2015 \\ \hline
\end{tabular}}

\par\end{centering}
\label{tab:RelatedWork}
\end{table*}

\subsection{Quantum Software Engineering}\label{Related:QSE}

QSE represents a recent and quantum-specific genre of the classical software engineering, providing practitioners with a systematic, process-oriented approach to developing software-intensive systems and applications for quantum computing \cite{X1_QSE} \cite{X5_QC}. Our review identified eight survey-based studies, summarised in Table \ref{tab:RelatedWork}, which help compare the scope and contributions of our proposed research. Recent surveys indicate that while traditional SE principles can be adapted for quantum software development, QC-specific challenges including, but not limited to operationalising QuBits/QuGates and quantum domain engineering, remain largely unaddressed. To derive new methods and processes, the QSE research community is striving to organise a body of knowledge and academic collaborations that can be stimulated via dedicated publication forums to disseminate quantum software specific research including, but not limited to IEEE QSW \footnote{IEEE International Conference on Quantum Software. \url{https://services.conferences.computer.org/2025/qsw/qsw-call-for-papers/}} and ACM TOSEM's Open Continuous Special Section on Quantum SE\footnote{ACM Transactions on Software Engineering and Methodology. \url{https://dl.acm.org/journal/tosem/quantumse}}. Academic discussions and published studies \cite{R1_ali2022software}\cite{R4_dyakonov2019will} highlight that beyond standard engineering principles, effective QSE requires a solid understanding of quantum mechanics \cite{R15_de2022software}. Many current software architects and developers, lacking this foundational knowledge, may be under-prepared for quantum software development \cite{X2_QSA}. Software architecture offers a way to abstract implementation complexities by representing systems as components and connectors. A recent systematic review identifies five key architecting activities to guide designers in incrementally engineering quantum software solutions \cite{X2_QSA}. Architectural solution(s) can also empower quantum software engineers to exploit architectural knowledge in terms of patterns (i.e., reuse knowledge), modeling notations (i.e., software representation), and tools (i.e., automation) to address the challenges of QSE. In addition to architecting, there is a need for empirically grounded guidelines and solutions to address the aspects of quantum software programming, model-driven development, and quality assurance in QSE \cite{x4_Issues}. 

Despite the recent advancements in QSE discipline, a multitude of technical and human skills related challenges persist that require innovative solutions and empirical research to address them. The study in \cite{QuantumFrontier2024} presents the results of a systematic mapping study of the current state of QSE research and identifies trends such as the most investigated topics, types and number of studies, main reported results, and the most studied quantum computing tools and frameworks. The study concludes by exploring the research community's interest in QSE by exploring \textit{how} the discipline is evolving, and \text{what} are the implications of research results as a source of information to facilitate knowledge transfer for researchers and practitioners. The authors of \cite{QPL2022} have conducted a two-phase empirical study to investigate the impcat of qunatum programming language on qunatum software development. Specifically, at the first stage, the authors mined data from GitHub repositories to investigate the quantum programming frameworks used by developers for qunatum software development. Secondly, the authors conducted a survey targeting the contributors of these repositories to capture their perspectives on the current adoption and challenges of quantum programming. The study \cite{QSEPractationers} conducted in-depth semi-structured interviews to investigate practationers' view on the issues and challenges of quantum software engineering. The insights from interviews provide developers’ perspectives such as the motivations, challenges, and outlooks within the rapidly evolving field of QSE. Despite promising advancements in QSE, significant challenges persist and demand attention for research and development. De Stefano et al. in \cite{QuantumFrontier2024} analysed widely-used quantum tools, offering valuable insights into their capabilities and limitations. Arias et al. in \cite{SECQSE} explored security issues in quantum programming languages and hardware, emphasising their impact on software development. The study highlighted critical vulnerabilities and advocated for the early integration of security measures to mitigate potential exploitation. Murillo et al. \cite{ChallengesQSE} further emphasised the need for robust frameworks and comprehensive security protocols. 

\subsection{Quantum Service Computing}\label{Related:QAAS}

Quantum service-orientation is an emerging trend driven by QC vendors, enabling developers to compose and invoke services on remotely hosted quantum systems \cite{R11_leymann2020quantum}. Table \ref{tab:RelatedWork} highlights four survey-based studies on quantum service computing. Platforms like Amazon Braket and IBM Q Experience have supported research on quantum cloud, quantum services, and QCaaS. Recent reviews highlight some prominent ongoing challenges, including limited hardware availability, quantum noise, and the quantum-classic split \cite{x7_QAAS}. Researchers are exploring ways to integrate microservices and quantum software practices, such as Quantum DevOps, to support the development of quantum microservices \cite{R23_khan2022agile}.

Researchers on quantum cloud computing and distributed quantum computing highlighted both significant advancements and ongoing challenges in leveraging quantum technologies for practical use. Both fields emphasised the importance of integrating quantum computing with existing infrastructures to expand computational capabilities and overcome current limitations. In this context, the study reported in \cite{R11_leymann2020quantum} provided a comprehensive review of quantum cloud computing, examining models, platforms, technologies, and critical issues such as resource management, security, and privacy. Their study identified open problems and outlined opportunities for future exploration. Similarly, the study \cite{x7_QAAS} explored the challenges and unresolved issues in distributed quantum computing, offering insights from a computing and communications perspective. Nguyen et al. \cite{QCCChallenge} provided a comprehensive review of quantum cloud computing, examining models, platforms, technologies, and critical issues such as resource management, security, and privacy. Their study identified open problems and outlined opportunities for future exploration. Similarly, Caleffi et al. \cite{DQC} explored the challenges and unresolved issues in distributed quantum computing, offering insights from a computing and communications perspective.


\section{Implications and Validity Threats}\label{sec:ImplicationsThreats}

This section discusses the potential implications of study results on research and practice of quantum computing as a service in Section \ref{subsec:implications}. We also present some threats to the validity of the findings from this study in Section \ref{subsec:threats}.

\subsection{Implications of Study Results}\label{subsec:implications}

This research focuses on investigating the potential synergy between quantum computing and quantum service computing in a rapidly emerging domain of quantum service computing, as illustrated in Figure \ref{fig:implications}. The study results synthesise software engineering principles applicable to QCaaS, defines service lifecycle phases in a reference architecture as a blueprint for quantum service-oriented development, and highlights key trends in terms of processes, professional expertise and tool support, etc. to guide future research and advance next-generation QCaaS solutions.  We highlight the implications of the study in terms of potential impacts of the results as valuable insights for both research and practice on quantum software engineering in general and quantum service computing in particular. 

\subsubsection{Implications for Research}
The results of this study are presented as structured catalogues (Table \ref{tab:DataExtraction}) and illustrative visualisations (Figure \ref{Fig-1:Context} - Figure \ref{fig5:RefArch}) that provide a synthesised knowledge on research state-of-the-art and vision for future needed research. Core findings of the study organised as service lifecycle activities in Section \ref{sec:RQ2SESolutions} and their integration into a reference architecture in Section \ref{sec:RefArch} can inform researchers about leveraging existing SOA principles, practices, and patterns in the context of emerging design challenges of QCaaS. Specifically, the reference architecture derived from existing literature provides a blue-print that harnesses patterns and tactics from classical software engineering (layered architecture, service facade) that can be integrated with quantum-specific patterns such as the quantum API gateway and quantum-classic split. An illustrative discussion of emerging trends can provide a roadmap as in Figure \ref{FutureResearch} for future research involving empiricism such as surveys, interviews, or repository analyses research to develop evidence-based taxonomies to architect and implement QCaaS. The study results also indicate that the `Manage' phase of the SOA lifecycle \cite{R20_keen2006patterns} is currently underexplored, offering potential for future research on quantum-specific service monitoring, governance, and quality management. Researchers could explore how LLMs might automate tasks such as generating quantum service models from natural language, completing quantum code, migrating code between frameworks, or simulating quantum circuits - speeding up the development via human-bot collaborative software engineering for QCaaS.

\subsubsection{Implications for Practice}
The results of this study provide evidence-based guidelines for practitioners, including quantum software architects and developers about the QSRs that guide quantum software and service development process (Section \ref{fig6:Implementation}). The pattern catalogue and thumbnails in Figure \ref{Fig:Pattern} represent a concentrated wisdom of quantum service that enables software designers and architects to rely on reuse knowledge and best practices to engineer QCaaS. The study indicates that QSE requires dedicated professional roles and expertise such as quantum domain engineer, quantum simulation manager, quantum algorithm desinger in an industrial scale development of unatum software as in Figure \ref{FutureResearch}.  The predominance of Python-based programming frameworks and qunatum service execution environments such as Qiskit and Amazon Braket reflects the gradual maturity quantum software/service ecosystem, reduces adoption risk, and enables practitioners to leverage established guidelines and concrete frameworks for service assembly and deployment. The study can help practitioners with the structured knowledge, patterns, and architectural blueprint needed to develop reliable and efficient quantum services for enabling and enhancing software engineering for QCaaS \cite{EASEGPT}.

\begin{figure}
    \centering
    \includegraphics[width=1.2\linewidth]{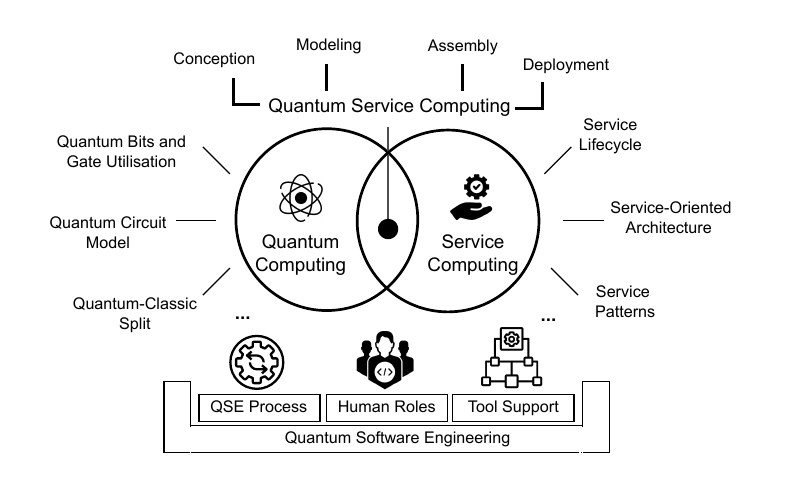}
    \caption{Synergising Quantum and Service Computing}
    \label{fig:implications}
\end{figure}

\subsection{Threats to Validity}\label{subsec:threats}

We now discuss several threats that may constrain or limit the validity of a study’s findings \cite{ValidityThreats2023}. These threats are commonly grouped into three main categories, as outlined below.

\subsubsection{Threats to Internal Validity} 

Internal validity addresses whether the results of the study (e.g., service lifecycle phases, and derived reference architecture) are influenced only by the variables under study (e.g., SMS process, ABD method). Potential threats to internal validity may arise from factors including, but are not limited to literature selection bias, the subjectivity of quality assessment criteria, and inconsistencies in data extraction that have been presented in Section \ref{ResearchMethod}. We tried to minimise these threats by following a structured process (Figure \ref{Fig-3:Method}), specifically customising the search string from Listing \ref{lst:searchquery} as per individual EDS by following the guidelines in \cite{EDS}. Moreover, to ensure a multistep quality assessment of each individual study, we derived a qulaity assessment checklist, as presented in Table \ref{tab:qualitycriteria}. Finally, to minimise the threat of omitting key information from the literature, we created a structured template as the data extraction form Appendix \ref{sec:appendix} - Table \ref{tab:data-schema} representing 11 data points to extract and record data from each primary study. 

\subsubsection{Threats to External Validity} External validity concerns the extent to which the study’s findings can be generalised beyond the immediate scope of this research, including different QC platforms, service models, or development environments. These threats can include the dominance of academic research over industrial perspectives/solutions, temporal restrictions of the reviewed literature, and the narrow evidence base, i.e., 41 studies as the underlying evidence for the proposed reference architecture. As the findings are derived primarily from peer-reviewed academic sources (Appendix \ref{sec:appendix} - Table \ref{tab:study-list}), they lack practitioners' perspective and industry-based insights for quantum service architecting and development. One possible way to minimise this threat is by using a mixed-method approach \cite{MMResearch} that uses findings from literature to derive an architecture and develops a proof-of-the-concept validation as in Figure \ref{fig6:Implementation}. The limitation can be further mitigated with future work that relies on survey research with quantum service developers and software engineers to validate and complement the findings of the study. Furthermore, extending the evidence base through approaches such as repository mining or practitioner-oriented evaluation of the reference architecture can provide stronger empirical grounding and enhance the generalisability of the proposed architecture.   

External validity could also be affected by study selection, i.e., publication bias. We adhered to the guidelines of systematic reviews \cite{SLRSLR} and only included the peer-reviewed published literature for data extrcation. By focusing primarily on peer-reviewed literature and excluding most gray or multivocal literature \cite{MLR} such as preprints, this study may underrepresent findings such as the service lifecycle phases and activities in each phase. While the decision to include only peer-reviewed published research enhances methodological rigour (per the guidelines of SLRs and/or SMS), it potentially limits the representativeness of available evidence as gray literature to limit the findings of this research.

\subsubsection{Threats to Construct Validity} Construct validity addresses the extent to which a study accurately measures what it is intended to measure, ensuring that the research questions, methodological details, etc. are clearly defined and appropriately operationalised. In the context of this study, construct validity relates to whether the selected research questions, study selection criteria, and data extraction items (in Section \ref{ResearchMethod}) adequately capture the essential aspects of QCaaS and its software engineering foundations. A key threat arises from the operationalisation of core constructs such as quantum service lifecycle, reference architecture, and QSRs. Although these constructs have been defined based on established software engineering and quantum computing literature \cite{X1_QSE} \cite{X2_QSA}, the rapid evolution of terminology within the quantum software engineering domain introduces a risk of conceptual drift, potentially limiting the generalisability of our findings to future research contexts.

Another potential threat is the subjectivity in the classification and coding of primary studies. Despite employing a structured data extraction form, interpreting constructs such as `pattern', `modelling notation', or `service quality attribute' in Table \ref{tab:DataExtraction} can be influenced by the authors’ judgment. To mitigate this threat, we adopted a multi-step verification process wherein the first author performed data extraction, and subsequent verification was independently conducted by co-authors.
\section{Conclusions}\label{sec:conclusions}

Quantum service-orientation enables QC vendors to provide end-users with access to computational resources—allowing `shots' on QPUs—through a utility computing model. Recent QSE research is combining service-orientation principles with quantum computing practices (algorithms, simulations, QuGates, etc.) to support QCaaS adoption by individuals and organisations, aiming for the anticipated leap into quantum computational supremacy. Using a systematic mapping study, we investigated (i) existing solutions for quantum service development, such as QSRs, modeling notations, patterns, and programming frameworks for architecting QCaaS, and (ii) emerging trends that highlight ongoing and future research needs. The results of this SMS as structured information (Table \ref{tab:DataExtraction}) and visual illustrations (Figure \ref{Fig-1:Context}, \ref{Fig:Results} - Figure \ref{fig5:RefArch}) highlight the strengths, limitations, and future research on QSE.  

The primary \textit{contributions} of this research focus on synthesising the existing research-based evidence on software engineering for QCaaS, proposes a reference architecture to guide quantum service lifecycle, and highlights emerging research trends on quantum service orientation. The \textit{findings} and \textit{implications} of the study can guide researchers with an evidence-based map of QCaaS and provide practitioners with patterns as reuse knowledge and architecture as a blue-print to implement solutions.

\textit{Needs for future work}:  The results of this SMS provide a foundation for future work along three main directions: (a) conducting a \textit{practitioners’ survey}, (b) \textit{validating the reference architecture} based on different case studies, and (c) \textit{mining social coding platforms} to strengthen empirical insights into QCaaS development. These findings can provided empirically-grounded foundations, employing mixed-method research approaches, thereby enriching the evidence base established through academic research and experimental analysis. Additionally, mining social coding platforms and open-source repositories (e.g., GitHub) will allow us to identify patterns, artifacts, and practices emerging from QCaaS developer communities.

\section*{Declaration of AI Assistance}
During the preparation of this article, the authors used AI assistance to refine the language, perform grammar checking, and improve sentence structure. After the AI assistance, the authors proof-read, edited, and thoroughly reviewed the content of the article, wherever required. The authors take responsibility for the contents presented in this article.

\printcredits

\bibliographystyle{elsarticle-num} 
\bibliography{References}

\balance

\appendix
\onecolumn
\section{Appendix}\label{sec:appendix}
\begin{lstlisting}[caption={QADL Script for Specification of Shor's Algorithm}, label={lst:QADLScript}]
@startqadl
Circuit ShorsAlgorithm {
    qubit q0, q1, q2
    gate Hadamard q0, q1, q2
    gate X q0   // Invert q0
    gate X q2   // Invert q2
    gate CNOT q0 q1  // Controlled q0, q1
    gate CNOT q1 q2  // Controlled q1, q2
    gate X q0   // Undo inversion of q0
    gate X q2   // Undo inversion of q2
    gate Hadamard q0, q1, q2
    gate X q0, q1, q2
    gate CNOT q0 q1  
    gate CNOT q1 q2
    gate X q0, q1, q2
    gate Hadamard q0, q1, q2
    measure q0 -> c0
    measure q1 -> c1
    measure q2 -> c2
}
@endqadl
\end{lstlisting}

\begin{table*}[h!]
\centering
\scriptsize
\caption{Data Extraction Template}
\label{tab:data-schema}
\renewcommand{\arraystretch}{1.15}
\setlength{\tabcolsep}{6pt}
\resizebox{\textwidth}{!}{%
\begin{tabular}{|>{\centering\arraybackslash}p{1.2cm}|p{5.2cm}|p{10.2cm}|}
\hline
\rowcolor{sectiongray}\multicolumn{3}{|c|}{\textbf{Generic Data}} \\ \hline
\textbf{D1} & \textbf{Study ID}      & The unique identifier for each study for its reference in the SMS \\ \hline
\textbf{D2} & \textbf{Study Title}   & The title of the study \\ \hline
\textbf{D3} & \textbf{Study Authors} & The authors as contributors of each study \\ \hline

\rowcolor{sectiongray}\multicolumn{3}{|c|}{\textbf{Demographic Data of Published Studies (RQ-1: Research Demography)}} \\ \hline
\textbf{D3} & \textbf{Year of publication}  & The specific year in which a study is published to identify the overall temporal distribution of publications \\ \hline
\textbf{D4} & \textbf{Type of publication}  & The type of research publications to identify publication foras such as conference proceedings or journal articles etc. \\ \hline
\textbf{D5} & \textbf{Type of Research Facet} & The type of research contribution to identify the strength of published evidence such as solution proposal, vision paper etc. \\ \hline
\textbf{D6} & \textbf{Application Domain}   & The type of system(s) or scenario(s) such as cyber security, smart city etc. that is supported by the proposed research \\ \hline

\rowcolor{sectiongray}\multicolumn{3}{|c|}{\textbf{Data for Quantum Service Development Lifecycle (RQ-2: Quantum Service Lifecycle)}} \\ \hline
\textbf{D7} & \textbf{Conception} & The data for conceptualising the quantum service \\ \hline
\emph{a} & \emph{Functionality} & Functional requirements as the conceived functionality offered by the service \\ \hline
\emph{b} & \emph{Quality}       & Quality attributes as the conceived quality of the service \\ \hline
\textbf{D8} & \textbf{Modeling}  & The data to model, i.e., specify or represent the conceived service \\ \hline
\emph{a} & \emph{Notation}       & Means to represent the design of the conceived service \\ \hline
\emph{b} & \emph{Pattern}        & Reusable knowledge or recurring solution(s) as applied best practices to design the service \\ \hline
\textbf{D9} & \textbf{Assembly}   & The data to assemble, i.e., program or implement the modeled service \\ \hline
\emph{a} & \emph{Usecases}        & The scenario that is implemented or automated by the assembled service \\ \hline
\emph{b} & \emph{Programming}     & Programming language or technologies used to assemble/implement the service \\ \hline
\textbf{D10} & \textbf{Deployment} & The data to deploy the assembled service \\ \hline
\emph{a} & \emph{Delivery}        & The quantum computing platform that is used to deploy and enable the execution of the deployed service \\ \hline

\rowcolor{sectiongray}\multicolumn{3}{|c|}{\textbf{Data for Emerging Trends and Potential Future Research}} \\ \hline
\textbf{D11} & \textbf{Emerging Trends} & Identify the emerging trends such as open challenges and/or needed solutions as potential dimensions of future research \\ \hline
\end{tabular}%
}
\end{table*}

\vspace{-0.5cm}

\renewcommand{\arraystretch}{1.15}
\setlength{\tabcolsep}{6pt}

\scriptsize
\begin{longtable}{|>{\centering\arraybackslash}p{1.2cm}|p{8.4cm}|p{5.4cm}|>{\centering\arraybackslash}p{1.0cm}|}
\caption{List of Primary Studies for the SMS\label{tab:study-list}}\\
\hline
\rowcolor{sectiongray}\textbf{Study ID} & \textbf{Study Title} & \textbf{Publication Venue} & \textbf{Year} \\ \hline
\endfirsthead

\hline
\rowcolor{sectiongray}\textbf{Study ID} & \textbf{Study Title} & \textbf{Publication Venue} & \textbf{Year} \\ \hline
\endhead

\hline
\multicolumn{4}{r}{\footnotesize Continued on next page} \\ \hline
\endfoot

\hline
\endlastfoot

S1  & Quantum software as a service through a quantum API gateway & IEEE Internet Computing & 2021 \\ \hline
S2  & QSOC: Quantum service oriented computing & Symp. and Summer School on Service-Oriented Computing (SummerSOC) & 2021 \\ \hline
S3  & QFaaS: A Serverless Function-as-a-Service Framework for Quantum Computing & Future Generation Computer Systems & 2024 \\ \hline
S4  & Quantum service-oriented computing: current landscape and challenges & Software Quality Journal & 2022 \\ \hline
S5  & Trials and tribulations of developing hybrid quantum-classical microservices systems & Quantum Software Engineering and Technology Workshop (QSWE) & 2021 \\ \hline
S6  & Towards Quantum algorithms-as-a-service & Int. Workshop on Quantum Programming for Software Engineering (Q-SE) & 2022 \\ \hline
S7  & Quantum Service-Oriented Architectures: From Hybrid Classical Approaches to Future Stand-Alone Solutions & Quantum Software Engineering & 2022 \\ \hline
S8  & Relevance of near-term quantum computing in the cloud: A humanities perspective & Int. Conf. on Cloud Computing and Services Science (CLOSER) & 2020 \\ \hline
S9  & Hybrid classical-quantum software services systems: Exploration of the rough edges & Int. Conf. on the Quality of Information and Communications Technology (QUATIC) & 2021 \\ \hline
S10 & A reference architecture for quantum computing as a service & Journal of King Saud University – Computer and Information Sciences & 2024 \\ \hline
S11 & Securing Smart Cities: Unraveling Quantum as a Service & Int. Workshop on Quantum Programming for Software Engineering (Q-SE) & 2023 \\ \hline
S12 & QCSHQD: Quantum computing as a service for Hybrid classical-quantum software development: A Vision & ACM Int. Workshop on Quantum Software Engineering (QSE-NE) & 2024 \\ \hline
S13 & A Design Framework for the Simulation of Distributed Quantum Computing & Workshop on High Performance and Quantum Computing Integration (HPQCI) & 2024 \\ \hline
S14 & Defining Quantum Advantage for Building a Sustainable MVP to Deliver Quantum Computing Services & Open Journal of Applied Sciences & 2024 \\ \hline
S15 & Development and Deployment of Quantum Services & Quantum Software: Aspects of Theory and System Design & 2024 \\ \hline
S16 & Integration of Classical and Quantum Services Using an Enterprise Service Bus & Int. Conf. on Product-Focused Software Process Improvement (PROFES) & 2023 \\ \hline
S17 & Enabling continuous deployment techniques for quantum services & Journal of Software: Practice and Experience & 2024 \\ \hline
S18 & Orchestration for quantum services: The power of load balancing across multiple service providers & Science of Computer Programming & 2024 \\ \hline
S19 & Qubernetes: Towards a unified cloud-native execution platform for hybrid classic-quantum computing & Information and Software Technology & 2024 \\ \hline
S20 & Quantum Service-Oriented Computing: A Practical Introduction to Quantum Web Services and Quantum Workflows & Int. Conf. on Web Engineering (ICWE) & 2024 \\ \hline
S21 & Quantum services generation and deployment process: a quality-oriented approach & Int. Conf. on the Quality of Information and Communications Technology (QUATIC) & 2023 \\ \hline
S22 & Quantum computing in the cloud: Analyzing job and machine characteristics & IEEE Int. Symposium on Workload Characterization (IISWC) & 2021 \\ \hline
S23 & Quantum as a Service Architecture for Security in a Smart City & Int. Conf. on the Quality of Information and Communications Technology (QUATIC) & 2023 \\ \hline
S24 & Quantum web services: Development and deployment & Int. Conf. on Web Engineering (ICWE) & 2023 \\ \hline
S25 & Quantum microservices: transforming software architecture with quantum computing & Int. Conf. on Advanced Information Networking and Applications (AINA) & 2024 \\ \hline
S26 & Evolution of Service-Oriented Computing: Integrating Quantum Techniques for Integer Factorization & IEEE Int. Conf. for Convergence in Technology (I2CT) & 2024 \\ \hline
S27 & Quantum and blockchain based Serverless edge computing: A vision, model, new trends and future directions & Internet Technology Letters & 2024 \\ \hline
S28 & Enabling continuous deployment techniques for quantum services & Journal of Software: Practice and Experience & 2024 \\ \hline
S29 & Quokka: a service ecosystem for workflow-based execution of variational quantum algorithms & Int. Conf. on Service-Oriented Computing (ICSOC) & 2022 \\ \hline
S30 & Quantum Cloud Computing from a User Perspective & Int. Conf. on Innovations for Community Services (I4CS) & 2023 \\ \hline
S31 & Towards a Quantum-Science Gateway: A Hybrid Reference Architecture Facilitating Quantum Computing Capabilities for Cloud Utilization & IEEE Access & 2023 \\ \hline
S32 & Hybrid Quantum Machine learning using Quantum Integrated Cloud Architecture (QICA) & Int. Conf. on Computing, Networking and Communications (ICNC) & 2023 \\ \hline
S33 & QuantoTrace: Quantum Error Correction as a Service for Robust Quantum Computing & Int. Conf. on Electrical Engineering and Information \& Communication Technology (ICEEICT) & 2024 \\ \hline
S34 & A Workflow for the Continuous Deployment of Quantum Services & IEEE Int. Conf. on Software Services Engineering (SSE) & 2023 \\ \hline
S35 & A quantum computing simulator scheme using MPI technology on cloud platform & IEEE Int. Conf. on Electrical Engineering, Big Data and Algorithms (EEBDA) & 2022 \\ \hline
S36 & Comparing the Orchestration of Quantum Applications on Hybrid Clouds & IEEE/ACM Int. Symp. on Cluster, Cloud and Internet Computing Workshops (CCGridW) & 2023 \\ \hline
S37 & Dynamic Quantum Network: from Quantum Data Centre to Quantum Cloud Computing & Optical Fiber Communication Conference (OFC) & 2022 \\ \hline
S38 & Stochastic Qubit Resource Allocation for Quantum Cloud Computing & IEEE/IFIP Network Operations and Management Symp. (NOMS) & 2023 \\ \hline
S39 & Architectural Vision for Quantum Computing in the Edge-Cloud Continuum & IEEE Int. Conf. on Quantum Software (QSW) & 2023 \\ \hline
S40 & A Quantum Safe Approach for Security Challenges at the Edge of Cloud in 5G and Beyond & IEEE Cloud Summit & 2024 \\ \hline
S41 & Quantum Software Architecture Blueprints for the Cloud: Overview and Application to Peer-2-Peer Energy Trading & IEEE Conf. on Technologies for Sustainability (SusTech) & 2023 \\ \hline

\end{longtable}

\end{document}